\documentclass[12pt,preprint]{aastex}

\newcommand{\psr}{\hbox{PSR~J0537$-$6910}}
\shortauthors{Middleditch et al.}
\shorttitle{Glitch Prediction in \psr}

\begin{document}

\title{Predicting the Starquakes in \psr
}

\author{J. Middleditch$^1$,
        F. E. Marshall$^2$, Q. D. Wang$^3$, 
        E. V. Gotthelf$^4$, and W. Zhang$^2$}

\altaffiltext{1}{Modeling, Algorithms, \& Informatics, CCS-3, MS B265,
                 Computer, Computational, \& Statistical Sciences Division,
                 Los Alamos National Laboratory, Los Alamos, NM 87545;
		 jon@lanl.gov}
\altaffiltext{2}{Laboratory for High Energy Astrophysics,
                 Goddard Space Flight Center, Greenbelt, MD 20771;
		 frank.marshall@gsfc.nasa.gov, William.W.Zhang@nasa.gov}
\altaffiltext{3}{Department of Astronomy, University of Massachusetts,
                 B-524, LGRT, Amherst, MA 01003; wqd@astro.umass.edu}
\altaffiltext{4}{Columbia Astrophysical Laboratory,
                 Columbia University,
                 550 West 120th Street,
                 New York, NY 10027; eric@astro.columbia.edu}

\email{jon@lanl.gov}

\begin{abstract}

We report the results of more than seven years of 
monitoring of \psr, the 16 ms pulsar in the 
Large Magellanic Cloud, using data acquired with the Rossi X-ray 
Timing Explorer. During this campaign the pulsar experienced 23 
sudden increases in frequency (``glitches'' -- 21 with increases 
of at least eight $\mu$Hz)
amounting to a total gain of over six parts per million of rotation 
frequency superposed on its gradual spindown of $\dot\nu = -2 \times 
10^{-10}$ Hz s$^{-1}$.  The time interval from one glitch to the next 
obeys a strong linear correlation to the amplitude of the first glitch, 
with a mean slope of about 400 days per part per million (6.5 days per
$\mu$Hz), such that these intervals can be predicted 
to within a few days, an accuracy which has never before been 
seen in any other pulsar.  There appears to be an upper limit
of $\sim$40 $\mu$Hz for the size of glitches in {\it all} pulsars,
with the 1999 April glitch of \psr\ as the largest so far.
The change of its spindown across the glitches,
$\Delta\dot\nu$, appears to have the same hard lower limit of 
-$1.5 \times 10^{-13}$ Hz s$^{-1}$, as, again, that observed 
in all other pulsars.  The spindown 
continues to increase in the long term, $\ddot\nu$ =
-10$^{-21}$ Hz s$^{-2}$, and thus the timing age 
of \psr\ ($-0.5 \nu {\dot\nu}^{-1}$) continues 
to decrease at a rate of nearly one year every year,
consistent with movement of its magnetic moment
away from its rotational axis by one radian every 
10,000 years, or about one meter per year.
\psr\ was likely to have been born as a nearly-aligned rotator
spinning at 75--80 Hz, with a $|\dot\nu|$ considerably smaller
than its current value of 2$\times$10$^{-10}$ Hz s$^{-1}$.  
Its pulse profile consists of a single pulse
which is found to be flat at its peak for at least 0.02 cycles.  
Glitch activity may grow exponentially with a
timescale of 170 years $\nu \dot\nu$ ${((\nu \dot\nu)_{Crab}})^{-1}$
in all young pulsars.
\end{abstract}


\section{Introduction}

The X-ray pulsar, \psr\ (hereafter J0537), in the 30 Doradus star 
formation region of the Large Magellanic Cloud (LMC), was discovered 
serendipitously in an observation of Supernova 1987A with the 
Rossi X-Ray Timing Explorer (RXTE) in a search for its pulsar 
remnant \citep{mar98,m00}.  Associated with the 4,000-year old 
LMC supernova remnant, N157B \citep{WG98}, with a rotation period 
of only 16.1 ms (62 Hz), it is the fastest rotating 
young pulsar known, the next fastest two young pulsars being the 
Crab pulsar, with a period of 33 ms (30 Hz), and PSR B0540-69, with a 
period of 50 ms (20 Hz).  With a spindown 
rate near $-2 \times 10^{-10}$ Hz s$^{-1}$, J0537 has one of the highest
energy loss rates known, of $\sim 5\times 10^{38} I_{45}$ ergs s$^{-1}$, 
where $I_{45}$ is the moment of inertia of the neutron star (NS) in units of
$10^{45}$ g cm$^2$, but it has so far only been detected in the 
X-ray band despite several radio and optical searches 
(Crawford et al. 1998; Mignani et al. 2000). 

Pulsars are known for their very stable rotation and 
small spindown rates, and their spindown histories can be 
used to constrain models of their emission mechanisms.
Characteristic pulsar ages, $\tau_c$, may be estimated, assuming 
that their spindowns are dominated by magnetic dipole 
radiation, $\dot\nu \propto \nu^3$, as $\tau_c \equiv 
-0.5 \nu {\dot\nu}^{-1}$, where $\nu$ is the rotation
rate in cycles s$^{-1}$ (Hz), and $\dot\nu$ is its (almost
always negative) time derivative, or spindown, in Hz s$^{-1}$.  
J0537 has a young characteristic timing age of $5\times 10^{3}$ 
years, consistent with the 4,000-yr estimated age of N157B,
though we have showed this quantity to be actually decreasing
with time (Marshall et al. 2004, hereafter M04).

Many pulsars have also been shown to undergo sudden 
discontinuities (usually increases) in their rotation rates 
known as ``glitches,'' (see, e.g., Alpar et al.~1993,1996; 
Hobbs et al.~2002; Johnston et al.~1995; Jones 2002; 
Lyne, Pritchard, \& Smith 1993; Lyne et al.~2000; 
Pines \& Alpar 1985;
Ruderman, Zhu, \& Chen 1998; Shemar \& Lyne 1996; 
Wang et al.~ 2000; Wang et al.~2001a; 
and Wong, Backer, \& Lyne 2001). 
Although glitches have been detected in both young and old 
pulsars, they are predominately found in the younger ones 
($\tau_c < 10^5$ yr, Urama and Okeke 1999, Lyne, Shemar, 
\& Smith 2000).  Glitches are thought to occur when 
angular momentum is transferred from a more rapidly rotating 
component of the NS to the outer crust (see also Anderson \& 
Itoh 1975; Franco, Link, \& Epstein 2000; and $\S$\ref{sec:nudisc}
for alternative models).  

The increase in 
rotation rate shows up within a few minutes because the NS star 
magnetic field is thought to be fixed in the crust.  Thus, in 
addition to the continuous spindown of a pulsar, glitches can 
be used to reveal details about the NS equation of state and 
internal structure, which would otherwise remain hidden (see, 
e.g., Datta \& Alpar 1993; Link, Epstein, \& Lattimer 1999).  
If, for example, erratic timing activity was found to 
always precede the large glitches, as seems likely from 
this work, it would help to eliminate mechanisms which rely 
on truly sudden onsets as plausible causes for glitching,
such as sudden events involving the unpinning of vast numbers 
of superfluid vortices within the crust, triggered by
mechanisms {\it other} than cracking in the solid crust.

Below we report ($\S$2) on more than seven years of observations 
of J0537, including more than four years of new observations, the 
whole of which argues that the time interval from one glitch
(the first) to the next glitch (the second) in J0537 is strongly 
correlated to the amplitude of the first glitch ($\S3$), a pattern 
that is similar to that of large quakes within the crust of our own 
planet.  In $\S$5 we compare the behavior of J0537 with that of the 
population of glitching pulsars and discuss how its glitch/time
correlation helps to discriminate between different proposed 
mechanisms triggering their glitches.  Section 6 lists our conclusions.

\section{Observations}
\label{sec:Obs}

The data were obtained with the Proportional Counter Array (PCA) on
board the RXTE observatory,
as described in detail in M04.  The PCA is sensitive to X-rays in
the 2--60 keV band and has moderate spectral resolution 
($\Delta E~E^{-1} \sim 18\%$).  Each event is time-tagged on the spacecraft
with an uncertainty of less than 50 $\mu$s \citep{Ro98}.
The background consists of unrejected charged particles and
X-rays from LMC X-1 and other cosmic X-ray sources.  Only
events in the first xenon layer in channels 5--50 ($\sim$3--20 keV)
were included in this study.  A log of RTXE observations 
of J0537 is given in Table 1.  The observations are separated 
by 23 glitches into two incomplete and 22 complete groups.
The number and epoch range for observations for which we have a 
timing solution (i.e., an
unambiguous cycle count linking its pulse time of arrival [TOA] to 
those of at least one neighbor) are listed under ``Phased,'' and the 
number and epoch range for the subset of these which were used to derive 
an ephemeris are listed under ``Fit.''  Our monitoring system of J0537 
continued as described in M04
except that, starting in late 2003 additional closely spaced observations 
were scheduled during the few week time period during which the next 
glitch was anticipated.  

\section{Data Analysis and Results}
\label{sec:DAR}

\subsection{The Pulse Profile}
\label{sec:PulsePr}

We first corrected the photon arrival events to the
Solar System Barycenter using the source position (J2000) of
$\alpha = 05^{hr}37^{m}47.36^{s}$,
$\delta = -69^{\circ}10^{\prime}20.4^{\prime\prime}$
\citep{W01b}.  Times are given in barycentric dynamical time (TDB).
Other details are as given in M04.
For each observation we determined the pulsar's frequency and
phase at the time of the first event
(a nominal spindown, $\dot\nu$, of -1.988$\times 10^{-10}$ Hz s$^{-1}$
was used throughout these processes).
The pulse shape was fitted at first to an ad hoc downward
parabola with a half width at zero height of 0.072 cycles,
but later to a master pulsar profile (MPP) derived from fitting 
folded data from many observations to a Breit-Wigner function
(see, e.g., Nelson et al.~1970).  
The uncertainty in phase was determined from Monte Carlo simulations 
to be $\sim$0.04 S/N$^{-1}$ cycles, where S/N is the signal-to-noise
ratio of the single peak of the folded J0537 pulse profile, including
noise fluctuations due to background and the pulsar signal itself.  
For the parameters of our observations (listed in M04) only about 
0.5\% of the count rate is due to the pulsar.  However, these events 
account for six times as much variance per event than does the 
background, due to their concentration within the pulse profile shape.

Although several more iterations of this fitting made the next 
MPP sharper, eventually this process actually resulted in lower S/Ns, 
at least until we utilized the frequencies as derived from the 
timing solutions (see \S3.2 below), which were much, much more 
accurate than could ever be achieved by fitting a single observation,
to derive improved phases from a simple, one dimensional 
Newton-Raphson fitting process.  The resulting MPP appeared to be 
wider at the top of the peak than any previously
derived MPPs, and could no longer be well fit by any Breit-Wigner
(BW) formula.  Accordingly, we subtracted a Gaussian 
of a given center, narrow width, and small amplitude, 
from the BW formula, and this fit our first timing solution 
MPP well.  Another MPP was generated by fitting to this modified
BW (MBW), and this appeared to be flat at the top for at least
0.02 cycles.  This master pulse profile and the
MBW used to produce it are plotted in Figure 1.

Further iterations, with MPPs generated using the 
MBWs made by fitting to the MPPs from the previous iteration, 
could still be well fit by an MBW, but also produced MPPs 
that were progressively more ``two-horned.''  These
MBWs also produced progressively worse S/Ns when fit to the 
individual runs, a possible side effect of the interaction of
statistics for the low S/N individual observations 
from which the MPP is produced,
with the maximization process for the fit response.
The MPP drawn in Fig.~1 is given by:
\begin{equation}
L(\phi) = {{1 - \phi (c_1 - \phi c_2 )} \over 
{1 - \phi (c_3 - \phi c_4 ) }}
e^{-c_5 {\phi}^2} (1 - c_6 e^{-c_7 (\phi - \psi)^2}),
\end{equation}
where $-0.5 \le \phi \le 0.5$.  
The three parameters that characterize
individual runs were determined by fitting a function of the form
$a L(\phi - \phi_0) + b$, where $a$ gives the signal amplitude,
$b$ the background, and $\phi_0$ defines the phase of the first
pulse arrival in the observation time interval.  
A time
of arrival (TOA) was obtained self-consistently by propagating the 
time of the first pulse peak arrival (derived from the measured pulse phase 
and the observation start time), by an integral number of cycles 
to some epoch near the center-of-gravity of the observation interval, 
using the nominal $\dot\nu$, together with the very accurate frequency 
given by the timing solution, or lacking that, the measured frequency 
for the individual observation.  The constants, $c_{1-7}$, 
$\psi$, $a$, $b$, and $\phi_0$ are given to full accuracy in Table 2. 
The phase, $\phi_0$, like $a$ and $b$, is in this case, for the MPP of 
Fig.~1, rather than any individual observation, and together with $\psi$, 
$c_{1-7}$, $a$, and $b$, makes a total of eleven parameters.

\subsection{The Timing Solutions}
\label{sec:Timesol}

The method of determining the timing solutions was
the same as has been frequently employed for pulsars and
other periodic phenomena.  Per M04, the initial time gap
between two observations ($\sim$6--12 hours) was only a few 
times longer than the total time spanned by each observation, 
(5,000--20,000 s), usually consisting of 2--4 contiguously
observed segments and their intervening gaps.   
This guaranteed that the average of 
the individual frequencies measured for two observations 
could reliably determine the mean frequency between the two. 
The time gap between the successive observations was then 
increased geometrically.  A value for $\dot\nu$, close to 
those already determined from fits to other segments, or, 
lacking that, the long term frequency history of J0537, was used
to count cycles over the longer time gaps.  If the estimate 
of the number of cycles between TOA's from two consecutive 
observations was within a few hundredths of an integer, and the
measured frequencies of the two observations were consistent
with the same frequency and the small correction for $\dot\nu$,
one could be confident that the cycle-counting was
unambiguous.  

Continued successes at cycle counting in extending
the baseline of the timing solution quickly reduced the
uncertainty of its validity to a vanishingly small value.  
When the time baseline of the observations and the gaps between 
them was sufficiently long, $\dot\nu$ could be reliably
determined by the fitting process.  Eventually, $\ddot\nu$ could
be determined when the time baseline exceeded 100 days.  The 
interval at which the gap size was reset to $\sim$6--12 hours 
was 60 days, and the whole timing procedure could be
started over, if necessary.  The cycle-counting process 
was extended over a longer and longer time baseline, until a glitch 
made further extension of the unambiguous cycle-counting impossible.  

A sum-squared minimizing routine was used to fit the data 
segments to a power law model, involving the pulsar braking 
index, $n$, in $\dot\nu \propto -\nu^n$.  
As an example, the frequency 
predicted by the power law relation at time t, can be, for 
the purposes of discussion, more conveniently expressed as: 
\begin{equation}
\nu (t) = \nu(t_0)( 1 - (\dot\nu(t_0)/\nu(t_0)) 
(n-1) (t - t_0) )^{-1/(n-1)},
\end{equation}
where $t_0$ is the epoch of the fit, and the quantity, 
$-\dot\nu(t_0) {\nu(t_0)}^{-1}$, is usually more concisely, 
though {\it not} as conveniently for the purposes
of discussion, expressed as $\gamma(t_0)$.
As a check on the power law parameters and their errors,
$\nu$, $\dot\nu$, and $\ddot\nu$, were derived, using 
simple linear algebra, from linear polynomial fits to 
the TOA's and cycle numbers of points belonging to the 
individual data segments: 
\begin{equation}
m_i = \phi_0 + \nu ({\rm TOA}_i-t_0) + \dot\nu ({\rm TOA}_i-t_0)^2/2
+ \ddot\nu ({\rm TOA}_i-t_0)^3/6,
\end{equation}
where $m_i$ is the integral number of cycles from time, 
$t_0$ - $\phi_0 \nu^{-1}$ to ${\rm TOA}_i$,
$\phi_0$ is the pulse phase at time, $t_0$,
and the need for $\phi_0$ can be eliminated 
by a proper choice of $t_0$.  
For another check, the power law fits were 
re-initialized with the equivalent 
parameters derived from the polynomial fits.
The difference between the 3rd order polynomial fits using 
$\nu$, $\dot\nu$, and $\ddot\nu$, and the power law fit 
that uses the equivalent parameters, 
consisting of terms of higher time
derivatives than $\ddot\nu$, was insignificant.  

In practice, when the data segments only contain
a few points, they are first fit
using only a fiducial phase or TOA, and $\nu$, with 
$ -\dot\nu \nu^{-1}$ and $n$ fixed, then with only
$n$ fixed, and then with all four parameters.
The parameters of the fits and their errors are listed 
in Table 3, along with an epoch of pulse peak arrival.  
In four cases of short data segments (5, 8, 13, and 17),
and three segments with $\ge$100\% errors on $n$ (1, 6, and 
11), $n$ was fixed at 3.5 without much increase in chi-square.  
In three more segments (10, 14, and 20) with large errors on 
$n$ and large chi-squared increases when $n$ was set to 3.5,
$n$ was fixed at values near to what the four-parameter 
fit would have produced (31.5, 35.0 and 19.0).  
Of this last three, segment 14 showed evidence that the
last point was high (see \ref{sec:TimeRes} and Figure 4).  
Segment 20 may have had both a small, but long-lasting 
recovery (the last and highest point of Segment 19 precedes 
it by only three days), and a later pre-glitch creep,
as its favored braking index near 19 can
only be diminished when observations were removed from 
near its beginning {\it and} its end, though more observation 
points would have been needed to be certain.  This situation 
is likely to become more common as further segments with more 
near-glitch observations are obtained.

\subsection{The Frequencies, and Spindowns}
\label{sec:nudotnu}

The errors on the frequency measurements of the individual
observations were also calibrated by Monte Carlo simulations,
and found to be [0.0418,0.045 cycles] (S/N T$_{rms}$)$^{-1}$, 
for the upper and lower 1$\sigma$ frequency errors, respectively.
Here, $T_{rms}$ is the square root of the mean squared moment in time 
for all of the events.  For a continuous (i.e., unbroken) observation 
of duration $T$, $T_{rms}$ is $(2 \sqrt{3})^{-1}~T$, as long as the 
signal and background count rates don't change with time.  
A simple way to understand these errors is that the phase error is the 
half width at half maximum, 0.044 cycles (Fig.~1), divided by
the S/N.  The frequency error is then that error divided by $T_{rms}$ 
(see, e.g., Ransom, Eikenberry, \& Middleditch 2002).  Figure 2
shows the individually measured frequencies of J0537 with a convenient 
$\dot\nu$ subtracted, along with the lines/curves representing the 
frequency trends from the timing solutions.  
 
The 23 glitches, ranging in size from 0.016 to 0.68 parts per million 
(ppm), are labeled in Fig.~2.  The mean rate of decrease, $-1.9760 
\times 10^{-10}$ Hz s$^{-1}$, is shown as the oblique dashed line.
All of the comments on the behavior of the glitching made in
M04 apply as well to this new total data set, now a factor of
three times larger.  The first glitch is still the largest
of the 23, and a measurement of the braking index, $n$, characteristic 
of the actual physical mechanism which slows the rotation of J0537, 
appears to be as elusive as ever.  However, the braking index of 
J0537 softens from 10.8 and 6.6 for the two halves of the long data 
segment 2, and there is some evidence for the same softening in most 
other long data intervals.  For the moment though, it is impossible 
to know if this is inconsistent with the behavior of the
high braking indices in non-glitching pulsars found by Johnston 
and Galloway (1999). 

The history of $\dot\nu$ is shown in Figure 3.  The points are
derived from timing solutions of subsegments of data with at 
least three observations, spanning at least 1,400,000 s, and 
with at least one observation no further away from the span midpoint 
than 35\% of the whole span.  Observations were appended to the 
subsegments until these criteria were met.  
A slight exception was made for data segment 17 (hourglass at 
$\sim$ MJD 53,140), which only spans two weeks.
Further observations from the end of the previous, fitted 
data subsegment were also prepended to the following subsegment 
when it would have otherwise spanned an insufficiently long time 
interval.  An extra subsegment was generated at 
the end of data segments 7, 19, and 20 by sharing the last 
observation of the (otherwise) last subsegment.

Figure 3 shows the same strong trend of $\dot\nu$ recovery (the
rapid change in $\dot\nu$, seen immediately after a glitch,
which diminishes with time) as has been sketched in Fig.~4 of M04.
Even in Fig.~2, the recovery of $\dot\nu$, in the long 
data section following the first glitch, is apparent 
as the decrease in the negative slope of the line representing the 
timing solution.  The long term trend, or steepening, from one line 
(timing solution frequency history) to those at later times, as first 
reported by M04, is also apparent in Fig.~2 by glitch 15.  

Any slope, including those between the subsegment $\dot\nu$'s in Fig.~3,
corresponds to a particular braking index $n$, as shown by the several 
dashed lines.  The slopes are in general very high at the beginnings of 
the data segments, and in general diminish somewhat by the end of the 
data segments.  The two parallel dotted lines are separated by 0.15 
pHz s$^{-1}$ and act as a guide for the changes of $\dot\nu$
within the data segments and the long term trend across all the
data segments, which, at -0.15 pHz s$^{-1}$ per five years, is an order
of magnitude smaller (and opposite in sign) than the typical recovery 
trend of $\dot\nu$.  Although the trends within the segments do not 
stay within the bounds delimited by the dotted lines, the {\it spans} of
the $\dot\nu$'s within each segment are approximately equal.
This will be discussed in further detail in section \ref{sec:fdott} below.

\subsection{The Timing Solution Phase Residuals}
\label{sec:TimeRes}

The phase residuals for the timing fits to all of the data segments 
are shown in Figure 4, plotted as in M04, except that 
we have included the phase recoveries of the third and 
fourth data segments, and most of recovery of the seventh data 
segment.  Doing this assures that the fits reflect 
the timing behavior of the pulsar during the segment as 
accurately as possible, and are best for measuring the changes 
in $\Delta\nu$ and $\Delta\dot\nu$ across the glitches 
(see $\S\S$\ref{sec:ft}--\ref{sec:fdott}).  
As far we or anyone else can determine, such
post-(large)-glitch recoveries in pulsars affect the timing 
in a continuous way, unlike the pre-glitch activity such
as that seen for data segments 7, 12, 19, and 21,
which will be discussed below.  We have not
plotted the 8 points\footnote{Note that one unphased point 
was plotted in both data segments 5 and 6, in Fig.~3 of M04 -- the 
point belongs to data segment 5 as can be seen in Fig~2 herein.}
for which there is no timing solution.

Other timing irregularities are also apparent 
in a half dozen other data segments shown
in Fig.~4, with both pre-glitch activity {\it and} 
a post-glitch recovery for data segment number 7 
(circles) -- the last and incomplete segment of M04 -- and in 
six cases have been omitted from the timing fits, and so are 
flagged by filling in the hollow characters.  Segment 7 is plotted 
with greater time resolution in Figure 5 (lower).  Conservatively, at least 
the first point (filled circle) is part of a recovery from glitch 6, 
which, at 0.46 ppm (M04), is still our second largest, and thus would 
have been expected if post- and pre-glitch activity scales with the 
size of the glitch.  Data segment six contains an unphased observation 
as its last point, supporting this hypothesis.  The points beyond MJD 
52,150 are shown in the inset frame of Fig.~5.  

In contrast, the microglitches at the end of data segments 12 and 21, plotted
in Fig.~4 as filled squares and diamonds, are very well fit by $\Delta\nu$'s
of 0.18 and 0.29 $\mu$Hz, or gains in frequency of about three and five 
parts per billion (ppb).  The duration of these features lasted for at least 
16 and 11, and possibly as long as 30 and 18 days.  This last interval
continued, starting with an unphaseable observation at MJD 53,695 (the 
last unphased point at the end of segment 21 in Fig.~2), followed by a 
glitch before the beginning of the next segment at MJD 53,702.  
Both data segments 14 (Fig.~4: bowties) and 19 (elongated diamonds) have 
last points which are several sigma above the trend.  Gaps of six and three 
days, respectively, follow these points.

In addition, segment 22 is also plotted in Fig.~5 (upper), and shows
evidence for quasi-stable behavior in the form of a steadily
increasing {\it rate} of ``creep'' in phase, broken by a {\it downward}
glitch, before resuming an upward trend.
The timing residuals for data segment 20 show more of a pure creep
without the downward point (solid pentagons in Fig.~4).
The creep  can be considered as part of an exponentially increasing
trend (see the Appendix), or as a slow creep of neutron superfluid
vortices in the crust (Link, Epstein, \& Baym 1993), or both.
The downward glitch behavior is rare, but a similar event apparently 
occurred within data segment 1, removed in time from any glitch 
(Fig.~4, leftmost squares).  It has now also occurred early in
data segment 23 (Fig.~4, rightmost squares).  We initially
speculated that glitch 23, expected near 2006 August
7.8 UT, was going to be a large one, but a glitch of
1.12 $\mu$Hz occurred around early UT August 4.  At the
time of this writing (late UT 2006 Aug.~25), the data from early UT 
August 21 was still clearly part of data segment 24 (see
Tables 1 \& 3, and Figs.~2, 4, \& 8), and glitch 24 is overdue.  

\subsection{The Glitch Amplitude-Time to Next glitch Correlation}
\label{sec:ft}

The glitches visible in Fig.~2 appear to have post-glitch,
non-glitching time intervals which are highly correlated to
their amplitudes.  In order to test this correlation
as accurately as possible, reliable measures of the glitch 
amplitudes and the time between glitches are desirable.
We have chosen to use the frequencies of the two observation points 
just preceding and following the glitch, as evaluated/extrapolated by/from 
each of the two timing solutions from the whole data segments which
preceded and followed the glitch, to determine the $\Delta\nu$ across 
the glitch.  This procedure yields two pairs of frequencies, one pair 
for the pre-glitch point from each the two timing solutions, which
yields the first estimate of $\Delta\nu$, and then another
similar pair for the post-glitch point, which yields the
second estimate of $\Delta\nu$, and the two estimates can
be compared as a check.  The agreement in all 23 cases is within a 
small fraction of one $\mu$Hz.  We then set the uncertainty on the 
time to the next glitch as a quarter of the sum of the gaps preceding 
and following the data segment between this glitch and the following.
Thus half the entire time range possible for a glitch is given by 
$\pm$ one error, and the whole range by twice this.

Table 4 lists the glitch amplitudes determined as described
above, while Figure 6 plots these amplitudes against
the time interval to the next glitch,\footnote{We
exclude from consideration the very small (ppb
in frequency) timing irregularities at the ends of the
data segments shown in Figs.~4 and 5}
for which the correlation coefficient is 0.94.  The 
coefficient against the post-glitch stable time interval,
i.e.~the maximum time interval following the glitch
when no glitches of any size are detected in the
extrapolation of the timing solution
(see the time interval correction in column seven of Table 
4), is 0.96 due to improvements to the fit of glitches
2, 3, 5, 6, 8, 11--13, 15, 17, and 20, with all of 
these points plotted in Figure 7.  Glitches 4, 9, 10,
14, 18, 21, and 22 were worse with the corrections, but 
only by small amounts each.  By contrast, the scatter 
of the amplitudes against the time to the previous glitch, 
is random, with a correlation coefficient of 0.05.

Figure 8 shows the integral of the glitch amplitude with
time.  The upper frame plots the actual - predicted 
glitch times.  The error bars are derived from the 
uncertainty in glitch onset times and, on the left hand
vertical axis, are converted into an 
error in $\mu$Hz through the slope of the oblique line drawn from 
the bottoms of glitch 1 to glitch 21 (but which was deliberately 
plotted offset by $+$10 days to allow easier comparison).  
The bottom frame of the figure shows occasional early glitching, 
such as the onset of glitch 3, which may persist early for several 
more glitches, evident as the match in slope between the lower 
corners and the oblique line.  Eventually an episode, during which 
only small (ppb) glitches occur, extends the time range of the data 
segment prior to the next large glitch, thereby resetting the glitch 
clock mechanism closer to what it was prior to the previous (early) 
glitch.  Data segments 7, 12, and 22, which follow glitches 6, 11, 
and 21, all of which end with quasi-stable behavior, appear to be such 
glitch-resetting segments.  Considerable variation about this clock 
must still be possible, as glitches 13--17 alternate late and early 
onsets.

\subsection{The Effect of Glitching on $\dot\nu$}
\label{sec:fdott}

The changes in $\dot\nu$ across the glitches, $\Delta\dot\nu$, 
and their errors, $\delta(\Delta\dot\nu)$, were determined in a 
fashion similar to those for $\Delta\nu$, i.e., by forward 
and backward extrapolation of the timing solutions, and
by evaluating $\dot\nu$ for the two observations bracketing
the glitch in time.  However, because the error in $\dot\nu$
is relatively much larger than that for $\nu$, and
depends strongly on the epoch(s) chosen for the 
timing solutions, it is necessary to get $\Delta\dot\nu$
directly from the timing solutions, with epochs set to
the times of the pair of observations which bracket each 
glitch (changing fit epochs was not necessary to measure the 
$\Delta\nu$'s).  The errors for $\Delta\dot\nu$ listed in Table 4 
are taken to be minimum of the two error estimates
from each of these pairs.

The pattern of the glitch recovery in $\dot\nu$ shown
in Fig.~3 indicates that the longer the interval which
precedes any given glitch, the smaller the values of
$|\dot\nu|$ and thus the larger the possible
$|\Delta\dot\nu|$ across the glitch.  However,
as mentioned in $\S$\ref{sec:nudotnu}, the post-glitch
$|\dot\nu|$ does not seem to be able to differ from
its pre-glitch counterpart by an indefinitely large amount.
These two facts suggest that there may be a correlation 
between the change in $\dot\nu$ across the glitch and the 
time interval from the {\it previous} glitch, and that it 
might saturate for high values.

The $\Delta\dot\nu$'s listed in Table 4 
(and plotted in Figure 9 against the duration of the 
previous intervals for 20 of the 23 glitches) are taken 
as the inverse-squared error mean of the two values of
each pair, whose difference also serves as a check of this 
procedure.  
The largest difference between these two is
50 ppm (0.05 on the abscissa of Fig.~9) for glitch 11.  
The next largest difference, for glitches 10 and 16, is only 
0.035 in Fig.~9, and all other differences are even
smaller.

For glitches with ${{1,000 \Delta\dot\nu} {\dot\nu}^{-1}} 
< 0.5$   (4, 5, 8, 10, 11, 13, \& 17) the correlation
coefficient is -0.88, while that for all glitches 
(2--22, excluding glitch 6 due to its large error) is only 
-0.77.  By contrast, the correlation of $\Delta\dot\nu {\dot\nu}^{-1}$
with the interval to the following glitch is 0.12 (i.e.,
a negative correlation {\it against} the predominantly negative
$\Delta\dot\nu$'s -- the same as with the 
correlation of 0.12 for  $\Delta\dot\nu$ vs $\Delta\nu$).
However, as suspected, the truly remarkable result 
is the hard upper limit of 0.075\% or 0.15 pHz s$^{-1}$ for the 
change.  The glitches near this limit, 2, 3, 7, 9, 15, 16, 18--21,
and 22, all have relatively long previous intervals, 
bearing out the suspected correlation, as also does 
glitch 12.  The correlation indicates 
that the interval which preceded the first glitch 
may not have been more than twice as long ($\sim$120 days) 
as the data segment for which observations have been taken, 
since the $\Delta\dot\nu$ across glitch 1 is 0.045\%.  It may also 
indicate that the true size of $\Delta\dot\nu {\dot\nu}^{-1}$ for 
glitch 14 was likely at the lower end of its error range,
near 0.4, rather than near the symbol plotted at 0.56 or higher,
in Fig.~9.  Noticeable improvements (a few to several units or 
so) in the chi-squares of the runs achieved, e.g., by using 
the polynomial fit parameters to initialize a minimum 
sum-squares routine, will typically result in changes to
$\Delta\dot\nu$ in Table 4 by only 1--2$\times$10$^{-14}$ 
Hz s$^{-1}$.  However, in none of these cases did the 
$\Delta\dot\nu$ ever exceed 0.15 pHz s$^{-1}$.

\section{Recapitulation}
\label{sec:Recap}

The observations of J0537 have revealed several unique 
and remarkable features ``in need of explanation.''
The first is the tight correlation 
between the size of the glitches ($\Delta\nu$) with the 
time interval to the following glitch, with a slope of
about 400 days ppm$^{-1}$.  The correlation with the post-glitch 
stable time interval is even tighter.  Second, the 
microglitches which precede the large glitches
can be due to a simple small increment to the pulse
frequency, $\nu$, or an initially gradual, but more than 
linear change in frequency with time, or a combination 
of both.  The data are also consistent with each glitch 
being followed by a smooth, but relatively small recovery 
involving the decay of only a few percent of the initial 
$\Delta\nu$. 

Third, the long term trend of $|\dot\nu|$ with time is {\it 
increasing}, and thus the timing age of J0537, 
-0.5$\nu \dot\nu^{-1}$,
continues to decrease at a rate of nearly
one year per year.  Another way of stating this is
that the gain in $|\dot\nu|$ across the glitch,
typically about 0.15 pHz s$^{-1}$, unlike the behavior
of $\dot\nu$ in Vela noted by Alpar (1998), is not 
completely given back before the following glitch, with 10 
percent of the gain remaining.  Fourth, the change in
$\dot\nu$, or $\Delta\dot\nu$, is correlated
to the time interval to the {\it preceding} 
glitch, but clearly saturates at 0.15 pHz s$^{-1}$.  
And finally, the RXTE-band ($\sim$2--20 keV) 
pulse profile has only a single pulse with a flat 
maximum for at least 0.02 cycles, or 320 $\mu$s.  

\section{Discussion}
\label{sec:Dis}

In $\S$\ref{sec:nudisc} we discuss the strong correlation
between glitch size and the time to the next glitch, and
explain how this helps in discriminating between 
current models applied to other glitching pulsars.  To this we add
a discussion of the change in spindown rate, $\Delta\dot\nu$,
across and between the glitches in
$\S$\ref{sec:dnudisc}, and a brief discussion on glitch latency
intervals in young pulsars in $\S$\ref{sec:young}.  The remaining 
longterm increase of $|\dot\nu|$, i.e., the increasing spindown 
with age, is then discussed in $\S$\ref{sec:Migrate}, including the 
possibility that the magnetic pole of J0537 is migrating away 
from its rotation axis, and the necessity for a glitch latency
interval in the youth of J0537 similar to that observed in the present 
day young pulsars.  Finally we finish this section with a brief 
discussion of the pulse profile, in $\S$\ref{sec:dispulsep}.

\subsection{The Glitch Size-Time to next Glitch Correlation}
\label{sec:nudisc}

It is generally accepted by now that the moment of 
inertia contained by the neutron superfluid within
the crust of an NS amounts to only about 1\% of the total,
which also includes both that of the superfluid interior 
and the solid crust.  
Some glitches, including those of the Crab pulsar and B0540-69, 
may be ``crustquakes,'' where the equilibrium configuration (EC) 
for the solid crust departs from its geometrical configuration as 
the pulsar spins down until eventually the crust cracks and settles.  

Of course, settling of the crust alone can not possibly 
be responsible for all glitches in Vela, as the NS would
run out of its supply of rotational oblateness after a
few hundred years.  Also, sufficient crust settling to 
produce a frequency change of only a fraction of a ppm, 
for a glitch in a low luminosity pulsar 
such as Vela, would involve the release of an
easily observable amount of energy (likely in the 
soft X-ray band), which was not seen \citep{AO95,Se00}.
In addition to the glitch models mentioned in this
work, several alternative models for glitching in pulsars
exist  (see, e.g., the discussion and references in $\S$6 
of Dall'Osso et al.~2003).  Other models have been proposed 
more recently (Chamel \& Carter 2006; Jahan-Miri 2006a,b; 
Negi 2006; Peng, Luo, \& Chou 2006; Zhou et al. 2004).  

In glitching pulsars, the solid crust of the NS 
spins down continuously in between glitches, as the 
magnetic field is locked to it.\footnote{Even now, it is not
yet clear how or if the interior superfluid of the NS is
coupled to the solid crust (Sedrakian \& Sedrakian 1995; 
Andersson, Comer, \& Prix 2004).}
During this time at least part of the neutron superfluid 
within the crust is decoupled from its spindown, and thus
becomes a second (and separate) stream.  Thus, as the
solid crust slows down, the vortex density of the second 
superfluid stream exceeds what would be appropriate for the 
solid crust (and the first superfluid stream) by an ever 
larger amount with time.  

Andersson et al.~(2004) have suggested that such a
two-stream superfluid system is subject to an instability,
analogous to the Kelvin-Helmholz instability, that acts
when the difference in effective rotation rates between the 
two streams is greater than a certain amount, $\omega_{cr}$.
During a glitch this causes a transfer of at least part of 
the extra angular
momentum in the second, and faster rotating superfluid stream,
to the solid crust and the first superfluid stream within it,
spinning up the pulsar by a certain rotational
difference, $\Delta\nu$.  Another view holds that
an excess of vortices from ``traps'' located within the crust, 
transfer angular momentum to it when triggered by a variety
of potential mechanisms, including crustquakes in addition to an
$\omega_{cr}$ (see, e.g., Alpar 1998; Alpar et al.~1993; 
Epstein \& Link 2000).  We explore this issue further below.

The actual mean size of a glitch in J0537 after about 10$^7$ s
(115.7 days) would only be about 18 $\mu$Hz, or just over 
one day of spindown (see glitch 3 of Table 4), which is 
nearly exactly one hundred times smaller
than the accumulated spindown over this time interval,
and thus supports the hypothesis of angular momentum transfer
from an NS component with 1\% of the moment of inertia
(399.1 days per glitching ppm for J0537 means that 0.9\%
of the spindown is being reversed by the glitching).  For the Vela 
pulsar,\footnote{Lyne et al. [2000] report 1.7\% for 
most pulsars.} the glitches average 2.0 ppm, or 22 $\mu$Hz, 
very close to the average glitch size for J0537, 
but this amounts to 17 days of spindown, or just over 
1.5\% of the accumulated spindown over the 2.8-year mean 
interval between its glitches, which again supports
angular momentum transfer from a 1\% NS component.  The difference
between 1.5--1.7\% in Vela and 1\% in J0537 may be due
to the difference in their Q values,\footnote{Q is the fraction 
of the $\Delta\nu$ gained across a glitch which decays afterward.}
as Wang et al.~(2000) report 0.38 for the 1996
October 13th glitch.

The coincidence between the mean absolute glitch sizes 
for J0537 and Vela suggests that the neutron superfluid in the 
crusts of {\it all} NSs,\footnote{The existence 
of a maximum in $\Delta\nu$ for the crustal
neutron superfluid also implies that the glitch of AXS 
J161730-505505 reported by Torii et al.~(2000) likely occurred 
closer to MJD 50,600 and had an amplitude near 26.2 $\mu$Hz, 
instead of between MJD 49,300 and MJD 50,000, when its amplitude 
would have had to have been between 42 and 115 $\mu$Hz.  We note 
that the timing age of AXS J161730-505505 is 16,000 years.}
when triggered by some mechanism,
which may not be the same for all pulsars,
favor dumping some of, 
or even more than (in the case of finite Q), $\omega_{cr}$.
Relatively ``large'' glitches, amounting 
to gains of 32.4 and 28.0 $\mu$Hz, still 
essentially within the range observed for both Vela and 
J0537, have also been observed from the 2 Hz PSR J1806-2125 
\citep{Ho02} and the 4 Hz PSR J1614-5048 \citep{W00}, 
respectively.  Although these increases in spin {\it frequency} 
are the same to within a factor of 2, the corresponding 
increases in energy span more than an order of magnitude, 
which lends yet more support for the storage and exchange 
of {\it angular momentum} by vortices in the neutron 
superfluid which is thought to lie within the crust.  

Although glitch 1 of J0537, at 42.2 $\mu$Hz, is the largest 
glitch ever observed, the 34.5 and 34.3 $\mu$Hz Vela 
glitches of 1978, July 13, and 2000, Jan.~16 come in a close 
second and third, followed by the 32.4 $\mu$Hz glitch from 
PSR J1806-2125 as a very close fourth, and still ahead of 
all of the other glitches seen from J0537 to date.  All known 
glitches with $\Delta\nu > 5~\mu$Hz, are listed\footnote{Zavlin, 
Pavlov, and Sanwal (2004) report three possible glitches in 
the 424 ms pulsar, 1E1207-5209, all with likely $\Delta\nu$'s 
$>$ 5 $\mu$Hz, but other interpretations for the timing 
irregularities also exist.} in Table 5, and histogrammed 
in Figure 10, which shows a clear maximum size for all glitches 
of 40 $\mu$Hz.  This lends further support for an $\omega_{cr}$, 
between the solid crust and part of the superfluid contained within 
it, and this maximum can not be much bigger than 40$\mu$Hz$\times$100, 
or 4 mHz, i.e., one ``revolution'' every 250 s \citep{AI75}.  This 
would correspond to 8 years of spindown for Vela, but only about 8 
months for J0537.  

The NS crust in Vela is nowhere near as stressed even 
after its 2.8-year mean inter-(large)glitch time interval
than it is in J0537 after only four months.  Specifically, the 
crust of Vela would have had to mostly settle by only 
0.007 cm after 2.8 years, whereas the crust of J0537 
would have had to mostly settle 0.06 cm in four 
months, as the settling distance, $\delta$R, where R is the 
radius of the neutron star and is assumed to be 12 
km,\footnote{The
mean settling rate [drop/time] goes as 
$4 \pi^2 \dot\nu \nu$R$^4$ (GM)$^{-1}$, and thus
would double for R = 14 km.} 
scales as $\nu\dot\nu$.  The factor of 70 difference in 
crust settling rates between the two may mean
that less than one large Vela glitch in 15 is initiated by crust 
cracking.  After J0537 and the Crab, the glitching pulsar with 
the next highest $\dot\nu\nu$ (down by a factor of 18) is the 15 
Hz J0205+6449.  The pulsar with the lowest $\dot\nu\nu$ (down by 
a factor of 12,000) is the 2 Hz J1806-2125, but its $\dot\nu$ is 
down only by a factor of 200.  Clearly most glitches in most 
pulsars can not be triggered by crust cracking; many of those in
the Crab and B0540, and almost all of those in J0537 likely 
are.\footnote{J0537 is so bright in most of the X-ray band, however, 
that the excess luminosity due to its crust settling might easily escape 
notice.  Assuming, in the case of J0537, some 10$^7$ s worth of 
spindown causes the settling of a 1.2$\times$10$^6$ cm radius by 
0.054 cm and a glitch of 17.8 $\mu$Hz, the specific energy gained
from the gravitational field of 1.3$\times$10$^{14}$ cm s$^{-2}$
would be 7$\times$10$^{12}$ ergs/gm.  Using a specific heat estimate
equal to that of water, or  4.17$\times$10$^{7}$ ergs gm$^{-1}$ K$^{-1}$, 
gives a temperature rise of 14.5 eV.  The temperature rise for
J0537's biggest glitch would be 34.5 eV.  With an NS surface 
temperature of 80 eV, these rises might not be detectable, unless 
the specific heat is overestimated by a large factor.
Link and Epstein (1996) have argued that such a medium deposition 
of energy within the NS crust could have caused the large glitch 
seen in Vela during 1988, Dec.~24 (Lyne, Smith, \& Pritchard 1992), 
and have also suggested the same mechanism for Crab glitches,
the difference in magnitude being due to the temperatures
of the two NS crusts, i.e., age.  One large Vela glitch out of the 
dozen known is consistent with the $>$15:1 ratio estimated above.
Unfortunately, no X-ray observations bracket the 1988 Vela glitch.}

Since the large glitches in J0537 are probably initiated by 
crust cracking events, the extreme linearity in its glitch 
size-post glitch quiet timing interval could be due to the 
steady divergence of rotation rates, $\nu_1$ and $\nu_2$, of the 
two superfluid streams,
and crust cracking occurring much more frequently than
its large glitches,\footnote{The cracking event glitch rate 
in Vela then would be larger than 1 in 15.}  
which only triggers them when $\nu_2 - \nu_1 > \omega_{cr}$.
The alternative, of having the amount of angular 
momentum dumped from many vortex traps proportional to
the amount of settling in its solid crust, is harder to justify,
as many vortex traps with differing
saturations would result in a chaotic glitch-time relationship.
The crust settling in J0537 would have to be a global, rather
than a local event, dumping the vast majority of traps to the same 
level, which seems unlikely. 

Another alternative would be to have just one trap \citep{FLE00,Jo02}, 
but the pre-(large)glitch microglitches visible in Figs.~4 and 5 
contraindicate this alternative.  
The relatively poor linearity observed in Vela glitches would then
result from two-stream rotational differences significantly
greater than $\omega_{cr}$ before cracking could trigger a glitch,
leakage due to vortex currents in the inter-glitch interval, or
both.  We note that a more complicated 
picture of the glitch processes within Vela is asymptotic to this 
linear relationship as the $\Delta\dot\nu$'s across its glitches 
become equal, which they very nearly are for the long term 
recovery components of the larger glitches of Vela 
\citep{Ap93}, certainly more so than for J0537.

Although it has been suggested that cracks play no role whatsoever 
in glitches (Jones 2003), the arguments are based on the difficulties 
of producing open cracks within the crust, while closed cracks are 
perfectly free to slide and grow, given sufficient stresses.
The crust does have a certain mechanical strength, 
even with cracks, but this can not literally hold it 
up as it strains against gravity for any indefinite period 
of time as the stresses continue to increase.  At stresses
greater than the yield stress,\footnote{For stresses below the
yield stress, a solid body undergoes a strain which is strictly
proportional to the applied stress, and this strain vanishes when
the stress is removed.  This is Hooke's Law for the elastic
behavior of solids.  At higher stresses the extra strain in the 
solid is permanent, and this is due to plasticity, crack growth, 
or a combination of both.  Another strain component which can
vanish when stresses are removed is due to crack opening.} 
the solid crust may have a range of higher stresses wherein it 
undergoes plastic deformation (Ruderman 1991a,b,c; Ruderman et al. 
1998).  In this stage the crust will slowly dissipate most of the 
gravitational energy involved in its settling, and thus avoid
a sudden, large {\it unobserved} release of gravitational energy
during a glitch (from whatever settling is left over).  
The older a pulsar is, the more it has glitched 
in the past, and thus the more extensive the network of cracks 
is likely to be.  

Eventually, for J0537, the stresses in the crust will 
exceed the stability criterion for a population 
of cracks which is always present, and these cracks 
then grow quickly in an unstable fashion (see, e.g., Dienes 
1985 and references therein) until they relieve the
stresses and subsequently stop growing, possibly coalescing 
into a fault which crosses the equator \citep{FLE00}, and
if $\nu_2 - \nu_1 > \omega_{cr}$, triggering a large glitch.
It is also possible that, during this process, the cracks 
will network throughout the entire crust as they grow and 
intersect each other.  
The mechanics of cracks in solid materials, including the 
NS solid crust, is discussed further in the Appendix.

A glitch can occur slightly early if
the previous glitch left cracks which were larger than
those which usually occur, and these become unstable
earlier (at lower stress levels), but would not cause a 
glitch which was too early because $\nu_2 - \nu_1 < \omega_{cr}$. 
In these cases the stress levels at which crack growth become 
unstable will be lower.  A glitch could also occur later if the 
population of larger cracks is relatively depleted.  

\subsection{The $\dot\nu$ Behavior}
\subsubsection{The $\Delta\dot\nu$ Behavior}
\label{sec:dnudisc}

We are still faced with the $\Delta\dot\nu$ across
the glitches, its correlation with the interval
{\it prior} to the glitch, the apparent saturation
value of 0.15 pHz s$^{-1}$, and its long term increase
in absolute value, in addition to the behavior of 
$\dot\nu$ between the glitches.  
If we consider that J0537, at 4,000 years of age, is a 
middle-aged pulsar which is much older than the Crab, 
and yet still less than half as old as Vela, the observed 
behavior of at least the $\dot\nu$ is what 
might be expected, at least empirically.  

The glitches in the Crab pulsar (and in B0540) are sufficiently small 
and infrequent, with a dozen glitches with $\Delta\nu \nu^{-1}$ from 
4$\times$10$^{-9}$ to 6$\times$10$^{-8}$ \citep{Wo01} between 1969 and 
1999,\footnote{Although the 1969 glitch of the Crab 
(MJD 40,493 -- Boynton et al.~1969) 
is generally quoted as 4$\times$10$^{-9}$ in $\Delta\nu \nu^{-1}$,
the observing coverage very near its time of occurrence was sparse,
and there is good reason to believe that the actual $\Delta\nu \nu^{-1}$
was a full 10$^{-8}$.  This is because a glitch was observed,
in data taken by Rem Stone for J.~E.~Nelson and J.~Middleditch, 
around 1 August, 1971 (very early in the [optical] observing season, 
which started earlier yet on 24 July), which was very accurately
measured to be 4.2$\times$10$^{-9}$ in $\Delta\nu \nu^{-1}$.
This glitch overlaid the 1969 glitch {\it nearly 
perfectly} when plotted on the same time scale but a 
different phase residual scale by a ratio of exactly 
2:5.  The timescale of an exponential decay of $\Delta\nu$ for 
this glitch, $\tau_c$, was  2.4 days, in good agreement with
$\tau_c = 3$ days reported by Wong et al.~(2001) for its
weaker glitches.} that they could only be due to a few vortex traps 
within the crust.  Thus the excess angular momentum dumped into 
the crust when a glitch occurs in the Crab is small, 
and the glitch size is small (Alpar 1998; Alpar et al.~1994,1996).  

In addition much of the gain in $\Delta\dot\nu {\dot\nu}^{-1}$ 
across its glitches, up to 4$\times$10$^{-4}$, persists
indefinitely, initially prompting some (Allen \& Horvath 1997; Link \& 
Epstein 1997; Link, Franco, \& Epstein 1998; Ruderman et al.~1998) 
to attribute this to a gain in the magnetic dipole moment due to 
an increase in angle between the magnetic and rotation axes 
(see $\S$\ref{sec:Migrate}).  However, the pulse profile of the Crab 
already has two peaks, and is thus hard to reconcile with pole migration.
Perhaps a more likely possibility, as first suggested by Alpar et al.~(1996),
is that the initial crackup permanently destroys the ability
of part of the Crab's solid crust to store vortices (creating ``vortex
depletion regions") thus reducing its effective moment of inertia,
which, for a constant braking torque, leads to a permanent increase
in spindown.  On the other hand, Ruderman (2005) attributes this 
effect to a permanent change of braking torque across the glitch 
due to an increase of the component of the dipole moment
perpendicular to the rotation axis, which, in turn,
is caused by migration of magnetic flux tubes and
vortices, rather than to any physical migration of the crust
along the NS surface.

In the much older (wrt the Crab) Vela pulsar, the NS solid 
crust may be saturated with cracks, allowing a considerable
number of traps, and/or a substantial second superfluid stream, 
so that the 
excess angular momentum dumped into the solid crust during 
a glitch spins the whole star up by an average near 22 $\mu$Hz 
every 2.8 years.  The extra $\Delta\dot\nu {\dot\nu}^{-1}$ 
gained across its glitches, 4--7$\times$10$^{-3}$, appears 
to completely die away in a what seems to be a linear fashion 
(i.e. with constant braking index) through vortex currents
\citep{Ap98}, and the subsequent glitch is primed to occur 
when this process finishes.  For Vela this means 
there's also a strong correlation between $\Delta\dot\nu$ 
across a glitch and the time to the next glitch, 
in addition to the same correlation for $\Delta\nu$
(Alpar et al.~1993).  

The situation for 
J0537 differs either because the crust frequently doesn't 
wait for currents to finish giving away all of the gain 
in $|\Delta\dot\nu|$ before causing a glitch, or because 
of an increase in the strength of the magnetic field, or
a small, permanent gain caused by realignment of the 
NS magnetic moment, or any combination of these.  
This does not require that the vortex currents in J0537 and 
Vela differ in any fundamental way, as the $|\Delta\dot\nu|$ 
gained from the glitch in J0537 subsequent to the early 
``giveaway'' glitch is usually correspondingly more modest, as 
can be seen in Fig.~3.

Although the glitches of the adolescent pulsar, J0537, average 
about six times smaller than those of Vela in $\Delta\nu \nu^{-1}$, 
they are an order of magnitude more frequent, so that its 
glitch activity parameter,\footnote{The glitch activity
parameter is defined as:
$A_g = \Sigma\Delta\nu_g / \nu /\Delta t$, 
where $\Sigma\Delta\nu_g$ is the sum of all spin frequency
gains to $\nu$ from glitches, over a timespan, $\Delta t$.}
is 0.9 ppm yr$^{-1}$ as compared to Vela's 
0.7 ppm yr$^{-1}$, and at 0.1--0.7$\times$10$^{-6}$,
they are still larger than those of the Crab by about one order of 
magnitude.  About 90\% of its gain in $\Delta\dot\nu  {\dot\nu}^{-1}$
(as opposed to 100\% in Vela), of up to 7.5$\times$10$^{-4}$ (about 
an order of magnitude smaller than the relative gains in Vela) 
``decays away'' prior to the next glitch, the only other difference
from Vela being the softening of the J0537 braking index between its
glitches.  Since the absolute size of the glitches in J0537 are just 
as large, and in one case larger than those in Vela, the crust of 
J0537 is likely just as saturated with cracks as Vela.  This 
may be more of a reflection of enhanced crust settling in J0537 due 
to its history of higher spin rates, than it is of actual age.

However, the relative amount of the gain in $\Delta\nu \nu^{-1}$ across 
a glitch in Vela that decays away afterward, or Q, at 0.38 \citep{W00} 
is about an order of magnitude larger than in J0537.
The integrated extra spindown 
due to a typical glitch of J0537 of size 
$\xi$ ppm, with a linear $\Delta\dot\nu$ recovery of 0.21 $\xi$ 
pHz s$^{-1}$ over the 400 $\xi$ days until the next glitch, would 
slowly remove 0.059 $\xi^2$ ppm from the gain in
$\nu$ across the glitch, during the interval to the next glitch,
which is consistent with the barely visible recovery 
following glitch 1 shown in Fig.~2.  

This decay may be caused by vortex currents from the 
interior superfluid to that in the crust, during the interval following 
a glitch, which makes up for an {\it excess} of vortices dumped 
from traps and/or a second superfluid during the glitch.  Interestingly, 
the agreement between 
the maximum values of the $\Delta\dot\nu$ of 0.15 pHz s$^{-1}$ involved 
in the long term recovery following the glitches of Vela, J0537, {\it and} 
the Crab (the 1989 July 13 glitch -- Wong, Backer, \& Lyne 2001) 
is nearly perfect, even better than that of the maximum values of 
$\Delta\nu$ among all pulsars. 
This is what would be expected from one NS to the next, as crusts, 
vortex currents and moments of inertia are basically the same,
perturbed possibly only by spin rate, magnetic field, and differences
in mass and temperature.  

\subsubsection{Young Pulsars and the Latent Interval}
\label{sec:young}

Livingstone, Kaspi, and Gavriil (2005) have called attention to the 
two orders magnitude difference in the glitch activity parameter, 
between the Crab pulsar and B0540, as evidence for significant
differences in the internal structure of neutron stars.  
Although the most obvious difference between these two is that 
between their pulse profiles, the harmonic content of B0540 (Seward, 
Harnden, \& Helfand 1984; Middleditch \& Pennypacker 1985) is 
similar to the optical pulse profile of Vela \citep{Wa77}, and 
distinct inclinations for the two sources could easily account for 
the rest.  It is more likely that young pulsars such
as the Crab, B0540, and B1509-58 don't glitch very 
frequently at first, having to rely on settling of their
crusts in order to form their initial cracks (generating vortex depletion
regions and vortex traps and/or more second superfluid domain in
the process), which then help in 
the further crackup of the NS crust.  In this case, the crustquake 
rate would increase exponentially (at least initially),
with the product, $\dot\nu \nu$, and the age of the pulsar, $\tau$: 
\begin{equation}
\nu A_g  \sim ~exp~({\tau \over \Upsilon}~ \dot\nu \nu / (\dot\nu
\nu)_{Crab})
\label{eq:nuAg}
\end{equation}
where $\Upsilon$ is an e-folding timescale of glitch activity. 
If we use the timing age of the pulsar, -0.5$\nu \dot\nu^{-1}$,
in place of $\tau$, then equation \ref{eq:nuAg} becomes:
\begin{equation}
\nu A_g \propto~exp~(- \nu^2 /(2 \Upsilon (\dot\nu \nu)_{Crab})~),
\label{eq:nu2Ag}
\end{equation}
and the e-folding time for B0540 would be three times as long 
as that of the Crab, due to its product, $\dot\nu \nu$, being three times
smaller than that of the Crab, plus 25\% because the timing 
age, $\tau$, of B0540 (1,550 years) is 25\% longer than that 
of the Crab (1,240 years).  

The absolute glitch activity, 
$\Sigma\Delta\nu_g/\Delta t$, for B0540 is really 75 times 
less\footnote{The Crab's $\nu A_g$ is taken from Livingstone et al.~(2005) 
to be 0.3 $\mu$Hz yr$^{-1}$, whereas that of B0540 is 0.004 $\mu$Hz yr$^{-1}$.} 
than that of the Crab, but this can be made up if $\Upsilon$ = 170 years.  
For B1509, with a 6.6 Hz spin, 
a 6.7$\times$10$^{-11}$ Hz s$^{-1}$ spindown, and a 1,670 
year timing age, its $\nu A_g$ is only 28\% of that of B0540,
consistent with no glitches having been observed to date from 
B1509, both pulsars having been discovered in the early 1980s.  
If no starquake is observed from B1509 in the next few 
decades, then perhaps equation \ref{eq:nuAg} will need a 
pre-exponential factor of $\dot\nu$, which is certainly needed 
for the glitching rate of mature pulsars like J0537 and Vela.

This is a simpler model than that employed by Alpar and Baykal (1994).  
The starquake rate must turn over and flatten out if the $\nu A_g$ 
predicted for J0537 is to match the actual value of 56 $\mu$Hz yr$^{-1}$, 
i.e., only a factor of 187 higher than that of the Crab.  This 
$\nu A_g$ is only a factor of 7 more than that of Vela, as compared 
to the factor of 12.5 expected from
the ratio of their $\dot\nu$'s, both pulsars being
old enough so that their cracking growth functions may have
reached the same high constant value.  This discrepancy is not
serious and can be completely resolved if the Q in Vela is 
about 0.46, a value close to that measured by some, or if
Vela is old enough to have a higher value of the cracking growth
function than that of J0537, or a combination of both.

\subsubsection{Magnetic Pole Migration}
\label{sec:Migrate}

Three facts support J0537 as being born, and possibly still 
remaining, close to an aligned rotator: the singly-peaked 
pulse profile; the low value of its $\dot\nu$ with respect to that 
of the Crab pulsar,\footnote{Although the two pulsars have 
identical spindown luminosities (-2$\pi^2 I \nu \dot\nu$) 
of 5$\times$10$^{38}$ ergs s$^{-1}$, close 
to the Eddington luminosity for an NS, SN 1986J, which is now 
producing 200 times the luminosity of the Crab Nebula in the 2 cm 
band (Bietenholz, Bartel, \& Rupen 2004), likely exceeds this in 
total luminosity by a large factor.  Thus the spindown of J0537 is 
not smaller than that of the Crab because its spindown luminosity,
and that of all pulsars, $L_{sd}$, is not likely to be limited
to $L_{edd}$, even by processes which we do not yet fully understand.
We also note that Shukre, Manchester, and Allen (1983) have suggested
that the SS 433 binary system contains an aligned, rotating,
accreting NS.}
which has almost twice the $\dot\nu$ and less than half
the $\nu$; and the long term increase of its $|\dot\nu|$.  The 
increase in the magnitude of the (negative) $\dot\nu$ amounts to about 
0.15 pHz s$^{-1}$ every five years, or 0.95$\times$10$^{-21}$ 
Hz s$^{-2}$, but is still uncertain by about 25\%.  Settling
of the NS over this time period due to spindown will produce 
a 0.1\% reduction in the 8.4 m NS equatorial, centrifugal bulge, or 
0.84 cm, and a change in the dipole moment due to scale size changes 
a thousand times smaller, if any at all because the magnetic axis 
is not close to either the rotational equator or pole.  

If the {\it entire} stellar moment of inertia can drop by
0.1\% over five years, then, provided the braking torque 
doesn't change, this could account for the 0.075\%
increase in $|\dot\nu|$ in the same period.  However,
if the effective magnetic dipole moment of the NS also drops,
the reduced braking torque could cancel the effect, but if it 
increases, it could help produce it (see, e.g., Ruderman 2005).  
If this effect is real
in J0537, then it is because its has the highest product,
$|\dot\nu \nu|$ of 1.2$\times$10$^{-8}$ Hz$^2$ s$^{-1}$.  However, 
the same product for the Crab pulsar, 1.1$\times$10$^{-8}$ Hz$^2$ 
s$^{-1}$, is a close 2nd, and the Crab clearly 
has a robust, positive $\ddot\nu$ of 1.2$\times$10$^{-20}$
Hz s$^{-2}$.  Still, some have suggested that the intrinsic
magnetic fields of pulsars may increase with time (see, e.g.,
Blandford, Applegate, \& Hernquist 1983), though
until now, there has been little evidence to support this
assertion \citep{W92}, and the magnetic field of the NS remnant 
of SN 1986J is clearly very strong after only 20 years 
\citep{Bie04}.  The effect in J0537 is consistent with a growth
in its magnetic field of 0.375$\times$10$^{-4}$ yr$^{-1}$, or 
35 MG yr$^{-1}$.

There remains only the migration of the magnetic axis as 
the possible cause of the long term increase of $|\dot\nu|$.
As mentioned in the previous section, some have initially argued 
that the persistent spindown gains in the Crab pulsar following 
its glitches are not caused by pole
migration (see, e.g., Link, Epstein, \& Baym 1992), but instead by 
the formation of new vortex depletion regions, \citep{Ap96}.
However, this only works for young pulsars
which haven't glitched at the high rates that Vela
and certainly J0537 have for many thousands of years.

Following arguments similar to those made previously by others
\citep{AH97,LE97}, and assuming that the spindown of J0537 is 
proportional to the effective magnetic moment, 
$|$\b m$_{eff}|
= |$\b m $\times$ \b {$\nu$}$| |$ \b {$\nu$}$|^{-1}$, 
squared, or $|$\b m$_{eff}|^2$, and that this quantity is
in turn proportional to $\sin^2{\alpha}$, where $\alpha$ is
the polar angle of \b m$_{eff}$ from the rotation axis,
and is also a linear function of time,
we have:
\begin{equation}
\alpha(t) = \dot\alpha t + \alpha_0,
\label{eq:alin}
\end{equation}
\begin{equation}
\dot\nu = \beta \sin^2\alpha(t)~({\nu(t) \over \nu_{psr} })^n,
\label{eq:powerlaw}
\end{equation}
\begin{equation}
\beta \sin^2{\alpha(\tau_{psr})} = -2 \times 10^{-10}~\rm{Hz~s}^{-1},
\label{eq:betanudot}
\end{equation}
where $n$ is the braking index (for now assumed to 
be 2.5 -- the same as is characteristic for the Crab pulsar),
$\dot\alpha$ is the time derivative of $\alpha(t)$,
$\tau_{psr}$ is the age of J0537, $\alpha_0$ is the 
initial value of $\alpha (t)$ for time 0, or $-\tau_{psr}$ from the present,  
and $\beta$ is a negative constant with dimensions of $\dot\nu$. 
As the magnetic pole migrates away from the rotation axis, the increase
in the effective magnetic moment produces the increase of the spindown as 
a function of time that is evident in Fig.~3.  Thus the time derivative 
of equation \ref{eq:powerlaw} is equal to the long term trend seen in 
Fig.~3:
\begin{equation}
\ddot\nu = 2 \dot\alpha \beta \sin{\alpha(\tau_{psr})} 
\cos{\alpha(\tau_{psr})} +  
n \dot\nu_{psr} \beta \sin^2{\alpha(\tau_{psr})} / \nu_{psr} 
 = -0.95 \times 10^{-21}~\rm{Hz~s}^{-2}.
\label{eq:betanuddot}
\end{equation}

We can remove the second term from the middle of equation \ref{eq:betanuddot}
by subtracting an assumed known present day power law contribution,
to $\ddot\nu$ for J0537, with $n = 2.5$, of $n \dot\nu_{psr} 
\beta \sin^2{\alpha(\tau_{psr})} / \nu_{psr} = n {\dot\nu^2}_{psr} 
/ \nu_{psr} = 1.6 \times 10^{-21}$ Hz s$^{-2}$, to get: 
\begin{equation}
\ddot\nu_{\alpha} = 2 \dot\alpha \beta \sin{\alpha(\tau_{psr})} 
\cos{\alpha(\tau_{psr})} = -2.55 \times 10^{-21}~\rm{Hz~s}^{-2}.
\label{eq:betanuddots}
\end{equation}
Dividing equation \ref{eq:betanudot} by equation \ref{eq:betanuddots} to
eliminate $\beta$ gives: 
\begin{equation}
{{\tan{\alpha(\tau_{psr})}} \over {2 \dot\alpha}} = 7.8 \times 10^{10}\rm{~s,~or} 
~2,475~\rm{years},
\label{eq:tan23}
\end{equation}
a transcendental equation which can be solved for $\dot\alpha$, with
$\alpha (t)$ as the simple linear function of time given in 
equation \ref{eq:alin}, and if $\alpha_0$ and $\tau_{psr}$ are 
known.\footnote{For a braking index, $n$, of only 2.0, near the 
lowest known
established, glitch-free braking index for any young pulsar
(Zhang et al.~2001; Cusumano, Massaro, \& Mineo 2003;
Livingstone et al.~2005), the right hand side of 
equation \ref{eq:tan23} increases to 2,830 years.}
Values for $\dot\alpha$ and $\alpha_0$ which make $\alpha(\tau_{psr})$ 
exceed 90 degrees are naturally excluded.  We can integrate
equation \ref{eq:powerlaw} to extrapolate backward and forward
to solve for $\nu(t)$ and $\dot\nu(t)$. 

Since the absolute value of the tangent function is always greater
than that of its argument, and $\dot\alpha$, and  $\alpha_0$ are, 
barring migration {\it toward} the rotation axis, positive, equation
\ref{eq:tan23} indicates that the age of J0537, $\tau_{psr}$, or at 
least, $\tau_{\alpha} \le \tau_{psr}$, the duration for which there 
is a constant $\dot\alpha$, can not exceed 5,000 years.  For 
reasonable values for $\alpha_0$ and $\dot\alpha$ with $\tau_{\alpha}$ 
= 4,000 years, equation \ref{eq:tan23} is a problem to solve.  For 
example, with $\alpha_0$ = 0.2 radians, no value of $\dot\alpha$ 
exists that gets the left hand side of equation \ref{eq:tan23} below 
3,368 years, as long as we assume $\tau_{\alpha}$ = 4,000 years.

The magnetic pole migration rate, necessary to render
equation \ref{eq:tan23} solvable, is plotted against its initial 
obliquity for various values of latency times, $\tau_{psr}$ 
- $\tau_{\alpha}$ for $\tau_{psr}$ = 4,000 years, in Figure 11 
(lower).  Four vertical lines for $\alpha_0$ = 0.1 to 0.4 radians,
in steps of 0.1 radians, intersect the lower portions of the 
curves for 22 (believable) values of $\dot\alpha$. 
From equation \ref{eq:alin} we have, for 
the present day magnetic obliquity, $\alpha_{psr}$:
\begin{equation}
\alpha_{psr} = \dot\alpha \tau_{\alpha} + \alpha_0 ,
\end{equation}
which we can rewrite for the duration of the active migrating
interval, $\tau_{\alpha}$ as:
\begin{equation}
\tau_{\alpha} = -\alpha_0/\dot\alpha + \alpha_{psr}/\dot\alpha .
\end{equation}
Thus, given $\alpha_{psr}$ and $\dot\alpha$, the relation between 
$\alpha_0$ and $\tau_{\alpha}$ is a straight line with a slope of 
$-1/\dot\alpha$ and an intercept of $\alpha_{psr}/\dot\alpha$.  This 
relation is plotted in the top frame of Fig.~11 for $\alpha_{psr}$ 
from 10 to 60 degrees in steps of 10 degrees, and illustrates those 
values of $\alpha_{psr}$ and $\alpha_0$ which are necessary to keep 
the non-glitching interval believably small.

The time histories of $\nu$(t) and $\dot\nu$(t), corresponding to the
pairs of ($\alpha_0$,$\dot\alpha$) from the lower frame of
Fig.~11, are plotted in Figure 12.  The figure shows a small 
range for the spin rate, $\nu$, of J0537 at birth,
between 75 and 80 Hz, and a much wider relative range for $\dot\nu$.  

\subsection{The Pulse Profile}
\label{sec:dispulsep}

The peak of the pulse profile of J0537 appears to be flat
for at least 300 $\mu$s (Fig.~1).  Following the line of argument
from Golden et al.~(2002) to test whether J0537 should
be able to produce high energy pulsations, given a
0.02 cycle extent of the plateau, 
we determine that the magnetic field of the emitting 
region, $B_{em}$ in Gauss is given by:
\begin{equation}
B_{em} = 10^{10} E,
\end{equation}
where $E$ is the energy of the peak of the pulsed
synchrotron emission in keV.  However, the Lorentz
$\gamma$ is only 8 from the same calculation,
so it is hard to know if this analysis is 
appropriate for J0537.  
Assuming that it is, and extending a dipole 
field of 0.925 tG, and a radius of 12 km,
gives 54 km for the location of pulsed keV
emission.  The field strength at a light cylinder with 
a 770 km radius would then be 3.5 MG, and thus
J0537 would be expected to produce optical pulsations.  
However, if J0537 is a nearly-aligned rotator,  the actual 
surface magnetic field may be near 3.5 tG, or higher.  In that 
case the pulsed keV emitting radius moves out to 86 km, and 
the magnetic field at the light cylinder increases to 14 MG. 

If the field does actually fall through 2--3 
$\times$10$^7$ G before reaching the light cylinder, by the 
logic of some (Cheng, Ho, \& Ruderman 1986a,b), one might 
expect an outer gap region where pulsed optical emission 
could arise.  Gil, Khechinashvili, and Melikidze (2001) model 
the pulsed optical emission from Geminga (Golden et al.~2002) 
assuming that, like J0537, it is a nearly aligned rotator.  
However, O'Connor, Golden, and Shearer (2005) argue that,
in the case of J0537, its pulsed radiation is synchrotron 
self-absorbed at optical wavelengths, with $m_V \sim 24$.  
Unfortunately, the limits on optical pulsation are not very 
stringent, with the magnitude 23.4 limit from Mignani et 
al.~(2000) derived from imaging alone and close to one solar 
luminosity, as compared to the Crab pulsar with 4 solar 
luminosities.

\section{Conclusion}
\label{sec:Conc}

The 62 Hz pulsar in the LMC, J0537, is unique among all
others.  It is the fastest spinning young pulsar, and the most 
actively glitching pulsar known, with a gain in frequency of 56
$\mu$Hz, as compared to Vela's 8 $\mu$Hz, yr$^{-1}$.
The extreme linearity of the glitch-size/post-glitch
stable time, in addition to other quasi-stable behavior, is 
consistent with the crack growth mechanism in the NS solid crust
for the initiation of its glitches, and a two superfluid stream
instability within the solid crust, as the cause of the variation 
in glitch magnitude.  Crack growth mechanisms may
also cause glitching activity in young pulsars, such as the
Crab and B0540, to increase exponentially with a 170-year timescale 
and the product, $\dot\nu \nu ({\dot\nu \nu_{Crab}})^{-1}$, and this 
may be detectable in the Crab after a few more decades.  Among all pulsars, 
the important quantity involved in glitches is {\it angular momentum}
transfer, i.e., $\Delta\nu$, rather than $\Delta\nu \nu^{-1}$ or
$\nu \Delta\nu$, and 40 $\mu$Hz appears to be an absolute maximum
for this quantity.  This represents an $\omega_{cr}$, or a
difference in rotation rate of one cycle per 250 s between the NS 
solid crust, and the superfluid vortices within it.  The moment 
of inertia of the crustal superfluid is close
1\% for J0537, lower than that ascribed to Vela and a few other
pulsars, and the difference may explain why Q, the fraction of
$\Delta\nu$ gained in the glitch which decays afterward, is an order 
of magnitude higher in these others. 

The 62 Hz rotation frequency of J0537 is over twice that of the 
30 Hz rate of the Crab pulsar, but yet its spindown rate is only 
slightly greater that half that of the Crab, and just barely 
larger than that of the 20 Hz rotator, PSR B0540-69.
Clearly its effective magnetic dipole moment is only about
25\% those of Vela, the Crab, and B0540.  It may just
have a weaker magnetic field than the others,
but its singly-peaked pulse profile argues that, when it was born
4,000 years ago, it was as a nearly-aligned rotator, spinning
at a rate between 75 and 80 Hz.  

The longterm increase in the magnitude of its (negative)
spindown, $|\ddot\nu|$,
near 10$^{-21}$ Hz s$^{-2}$ (or 2.55$\times$10$^{-21}$
Hz s$^{-2}$ after correcting for braking by magnetic dipole 
radiation and/or pulsar wind processes), supports the
interpretation of magnetic pole migration away from the
rotation axis over time, by about one radian every 10,000
years, or one meter on the NS surface per year, or even less.  
This rate agrees with the estimate, made in the Appendix, of the 
crack growth rate necessary to trigger the glitches in J0537.
Like the Crab, or B0540, which, after 950 and 1,550 years, 
are both still in a low glitch-activity state, the interval of 
relatively low glitch activity in J0537 may have lasted
1,500 years, although realistic values for the initial
obliquity and migration rate also exist for an interval as
short as 400 years.  
The spindown rate, $|\dot\nu|$, for J0537 at birth was 
likely at least 30\% less than it is today.

There also appears to be a maximum absolute value for the 
(almost always negative) change in spindown, $|\Delta\dot\nu|$, 
across (and for J0537 and Vela also equal to that involved in 
the longterm recovery following), a glitch, 
in all pulsars, of 0.15 pHz s$^{-1}$ (i.e., a {\it minimum}
value of -0.15 pHz s$^{-1}$).  This is equivalent 
to a maximum value for vortex currents within pulsars, though 
exactly how or why this occurs is still unknown.
Finally, the single peak of the J0537 pulse profile is flat at its top
for at least 0.02 cycles, or 320 $\mu$s.

\acknowledgments J.M.~thanks Drs.~Ali Alpar, John Dienes, Aaron Golden, 
Naoki Itoh, David Pines, and Mal Ruderman for helpful, guiding 
conversations, and his group, CCS-3, for support during a substantial
interval when he was without funding.  E.V.G.'s research is supported by 
NASA grant NAG05GR11G.  This research has made use of data obtained 
through the High Energy Astrophysics Science Archive Research Center 
Online Service, provided by the NASA/Goddard Space Flight Center.  This 
research was performed in part under the auspices of the Department of 
Energy.

\section{Appendix}
\appendix

From Dienes (1983) as
corrected by Rice (1984), a penny-shaped crack will 
become unstable to rapid growth if the following
inequality holds:\footnote{Cracks in the shape of a disk lying 
in a plane are the only ones, aside from elliptical cracks, 
for which there is a known stress solution.}
\begin{equation}
(\sigma - \tau)^2 c > {\pi \over 2}
{{2 - \nu_p} \over {1 - \nu_p}} \gamma_{se} \mu ,
\label{eq:uns}
\end{equation}
where $c$ is the crack radius, $\sigma$ and $\tau$ are 
the traction and interfacial friction across the crack,
$\nu_p$ is Poisson's ratio\footnote{A cylindrical rod of height,
$H$, which undergoes a given small ${\delta H} H^{-1}$ when subjected 
to a uniaxial stress, will increase in radius by
a ${{\delta R}/R} = \nu_p \delta H  H^{-1}$.  Poisson's ratio
for an incompressible material is 0.5.}, $\gamma_{se}$ is the 
specific surface energy (in, e.g., ergs cm$^{-2}$, 
or dynes cm$^{-1}$),\footnote{The $\gamma_{se}$ 
for water is 77 ergs cm$^{-2}$.  Most familiar solid
materials have $\gamma_{se}$'s well into the thousands.  However,
the $\gamma_{se}$ for HMX crystals is only 50!}  
and $\mu$ is the shear modulus.\footnote{If a solid body
of height, $H$, is subjected to a small shear stress in an
orthogonal direction to $H$ by a pressure of $P$ dynes cm$^{-2}$,
and shears by $\delta L$ in that direction, then its
shear modulus, $\mu = P H (\delta L)^{-1}$.}
As the crack radius, $c$, 
increases, unstable growth will occur for lower stress levels.

Melting and ``healing'' of the cracks due to interfacial 
friction may also occur as their two surfaces slide against 
each other.  Others may stop growing when they intersect 
neighboring cracks at a high angle (``T'' cracks).  Still 
others may rotate or translate within the solid crust so 
that the traction stress is reduced across the largest and 
hence, most unstable cracks, and others yet may disappear 
through subduction below the crust
\citep{R91a}.  As occurs with faults in the Earth's crust 
(see, e.g., Scholz 2002), the crack intersections in the NS 
crust and other local variations prevent the crustquake process 
from relieving {\it all} strains necessary for it to settle to 
its current EC.  

As stated in the main text, data segments 7, 12, and 21, 
which may restore the ``original'' glitch timing shown in 
Fig.~8, all have a period of quasi-stable behavior during this
interval, instead of a prompt (large) glitch.  This behavior
can be physically interpreted as the NS still having
the ``guilty'' cracks which caused the onset of the 
previous glitch, except that these have been partially
``repaired'' by melting or some other mode of
fusion for closed cracks.  The repair process
leaves the crack radius smaller, thus revising
the failure stress upward toward the nominal value.  
As mentioned above, another possibility is
rotation or translation of the crack over several consecutive
glitch intervals until the traction stress at the time of
early onset is no longer sufficient to cause the crack to grow 
in an unstable manner.

For the cases such as the end of data segments 20 and 22 (Figs.~4
and 5) 
which shows a very gradual onset of angular momentum transfer,
it is necessary to show that crack growth can have a 
gradual, slow rise in velocity from zero, or a ``creep'' 
phase.  Evans (1974) and Charles (1958) have observed that,
when the stress is below the critical level of
instability given above in Equation (1), cracks may
grow at low speeds:
\begin{equation}
v = v_R ({K/{K_1}})^k,
\end{equation}
where $v$ is the crack growth speed, $v_R$ is the 
Rayleigh velocity, $0.93 \sqrt{\mu   \rho^{-1}}$,
the speed of a surface (shear) wave in the material of
density, $\rho$, $K$ is the stress intensity factor at the 
crack tip,
$\sqrt{\pi c} (\sigma - \tau)$, $K_1$ is some stress
intensity scale factor, and the optimum $k$ is 12
for many brittle materials.  The gradual behavior 
observed in data segments 20 and 22 (Fig.~5) may
be the result of such a process.  This formula holds 
only for a stress intensity factor, $K$, below a certain
critical value, $K_0 = \sqrt{2 \pi E \gamma_{se}}$, where
$E$ is Young's Modulus\footnote{A body of length, $L$, when
subjected to a small tensile force per unit area, $\sigma$, 
will strain by an amount, $\delta L = \sigma L E^{-1}$, where $E$ 
is the Young's Modulus (and $\delta L$ will drop to 0 when 
$\sigma = 0$).  The shear modulus, $\mu$ is related to $E$ and 
Poisson's ratio, $\nu_P$, by $E = 2 \mu (1 + \nu_P)$.}, and 
$\gamma_{se}$ again, is the specific surface energy.  For higher 
stress concentration factors, Freund (1990) has shown in 
theoretical studies that:
\begin{equation}
v = v_R (1 - ({K_0/K})^2),
\end{equation}
and this equation may likely govern the crack speed
behavior for all sudden glitches.  A value for the scale 
of the stress concentration factor, $K_1$, can be 
derived by setting the values and slopes of the $v$'s 
for the two regions equal:
\begin{equation}
K_1 = K_0 (k/2)^{1/k} (1 + 2/k)^{1/2 + 1/k} .
\end{equation}
From equation \ref{eq:uns}, the critical radius for closed 
penny-shaped cracks is:
\begin{equation}
c_{cr} = {\pi \over 2} {(2 - \nu_P) \over (1 - \nu_P)} {{\gamma_{se}
\mu} \over (\sigma - \tau)^2},
\label{eq:ccr}
\end{equation}
and the corresponding fracture toughness is then:
\begin{equation}
K_0 = \sqrt{\pi c_{cr}} (\sigma - \tau) = \pi \sqrt{(1 - \nu / 2)
\gamma_{se} \mu (1 - \nu)}.
\end{equation}
And finally, the relative crack speed at which
the two stress concentration factors are equal is:
\begin{equation}
v/v_R = 1/(1 + k/2),
\end{equation}
which is 1/7 for $k = 12$.

We can estimate\footnote{Or calculate, as is done in Strohmayer et 
al.~(1991)} the critical crack radius from equation \ref{eq:ccr}
by taking $\mu$ = 2$\times$10$^{29}$ dynes cm$^{-2}$ from Ruderman 
(1991b), the values for the displacement field near 0.05 cm
from Baym and Pines (1971 -- also as equations 1 \& 2 in Link et 
al.~1998), a $\nu_P$ = 0.5 for incompressible matter, and amortizing 
this displacement over 1 km of crust to get a strain near 
5$\times$10$^{-7}$.  Using the strain times a Young's modulus 
of 3$\mu$ = 6$\times$10$^{29}$ dynes cm$^{-2}$, yields
a stress of 3$\times$10$^{23}$ dynes cm$^{-2}$, or
3$\times$10$^{14}$ kBar, reassuringly close to yield stresses for 
normal, everyday metals scaled up to NS mid-crustal densities.  
For a realistic $\gamma_{se}$, we can scale up 10$^4$ ergs cm$^{-2}$, 
for a material with a mean density of 5, by 1$\times$10$^{13}$ to 
match the density of the outer crust, and we can ignore, for the 
moment, the interfacial friction, $\tau$.  Using equation \ref{eq:ccr} 
the resulting critical crack radius, for the onset of
instability for these (hypothetical) conditions in the
crust, is c$_{cr}$ = 5.5 cm.  Of course this value will increase
for any non-zero value for the interfacial friction, with
a $\tau$ of half of the 3$\times$10$^{14}$ kBar stress 
given above increasing c$_{cr}$ to 22 cm.  Thus the total growth
needed per year for the glitches then approaches one meter,
in agreement with the amount of pole migration required to
account for the persistent increasing trend in $|\dot\nu|$.

The crust will also have some distribution of
crack sizes, as occurs in the Earth's crust, which
is exponential over many orders
of magnitude \citep{Sz02}.  In this case, there
will be some probability that a crack of
a certain size exceeding some limit will be
present in the volume of interest.  If the
volume is not large enough, then there
might well be no crack larger than that limit,
and thus statistics enter into calculations
of NS crust failure, just as it does in calculations
of mechanical failure for other solid materials.


\begin{deluxetable}{cccccccccc}
\rotate
\tablewidth{0.0pt}
\tablecaption{Observing Log}
\tablehead{
   \colhead{Segment} &\colhead{Total number of} & \colhead{Start-End} 
& \colhead{Number of Phased} & \colhead{Start-End}
& \colhead{Number of fit} & \colhead{Start--End\tablenotemark{a}}  \\
 \colhead{Number} & \colhead{Observations} &  \colhead{MJD-MJD} 
& \colhead{Observations} &
\colhead{MJD-MJD} &\colhead{Observations} & \colhead{MJD - MJD} \\
  }
  \startdata
   1 & 11 & 51197 - 51276 & 10 & 51197 - 51262 & 10 & 51197 - 51262 \\
   2 & 30 & 51294 - 51560 & 28 & 51310 - 51546 & 28 & 51310 - 51546 \\
   3 & 17 & 51576 - 51705 & 17 & 51576 - 51705 & 17 & 51576 - 51705 \\
   4 & 18 & 51715 - 51818 & 18 & 51715 - 51818 & 18 & 51715 - 51818 \\
   5 &  9 & 51833 - 51874 &  8 & 51833 - 51864 &  8 & 51833 - 51864 \\
   6 & 10 & 51886 - 51954 &  9 & 51886 - 51941 &  9 & 51886 - 51941 \\
   7 & 31 & 51964 - 52165 & 31 & 51964 - 52165 & 25 & 51973 - 52144 \\
   8 &  9 & 52175 - 52229 &  9 & 52175 - 52229 &  9 & 52175 - 52229 \\
   9 & 22 & 52252 - 52382 & 22 & 52252 - 52382 & 22 & 52252 - 52382 \\
  10 & 11 & 52389 - 52445 & 11 & 52389 - 52445 & 11 & 52389 - 52445 \\
  11 & 13 & 52460 - 52539 & 13 & 52460 - 52539 & 13 & 52460 - 52539 \\
  12 & 31 & 52551 - 52733 & 31 & 52551 - 52733 & 27 & 52551 - 52715 \\
  13 & 10 & 52745 - 52814 &  8 & 52745 - 52792 &  8 & 52745 - 52792 \\
  14 & 12 & 52822 - 52884 & 12 & 52822 - 52884 & 12 & 52822 - 52884 \\
  15 & 12 & 52889 - 53007 & 12 & 52889 - 53007 & 12 & 52889 - 53007 \\
  16 & 14 & 53019 - 53122 & 14 & 53019 - 53122 & 14 & 53019 - 53122 \\
  17 &  5 & 53128 - 53142 &  5 & 53128 - 53142 &  5 & 53128 - 53142 \\
  18 & 15 & 53147 - 53285 & 15 & 53147 - 53285 & 15 & 53147 - 53285 \\
\\
\\
\\
  19 & 21 & 53290 - 53443 & 21 & 53290 - 53443 & 20 & 53291 - 53440 \\
  20 & 15 & 53446 - 53549 & 15 & 53446 - 53549 & 11 & 53446 - 53538 \\
  21 & 20 & 53551 - 53696 & 19 & 53551 - 53696 & 16 & 53551 - 53682 \\
  22 & 19 & 53711 - 53859 & 19 & 53711 - 53859 & 14 & 53711 - 53839 \\
  23 & 13 & 53862 - 53950 & 13 & 53862 - 53950 & 10 & 53862 - 53933 \\
  24 &  4 & 53953 - 53968 &  4 & 53953 - 53968 &  4 & 53953 - 53968 \\
  \enddata
\tablenotetext{a}{Observations fit to determine the pulsar ephemerides
  given in Table 3.}
\label{tab:obstab}
\end{deluxetable}

\tablehead{
\ &  & \colhead{All: Start-End} & & \colhead{Phased: Start-End}
& & \colhead{Fit: Start--End\tablenotemark{a}}  \\
\colhead{Group} & \colhead{Number} &  \colhead{MDJ-MJD} & \colhead{Number} &
\colhead{MJD-MJD} &\colhead{Number} & \colhead{MJD - MJD} \\
  }

\begin{deluxetable}{cccccrl}
\small
\tablewidth{0.0pt}
\tablecaption{Master Pulse Profile (MPP)\tablenotemark{a,b} }
\tablehead{
   \colhead{Constant Name} &&&&& \colhead{Value} \hfil \\
  }
  \startdata
   $a$        &&&& &~~~98,969.&6       \\
   $b$        &&&& &1,791,946.&      \\
   $\phi_0$   &&&& &~~~~~~~~0.&5050471985806    \\
   $\psi$    &&&&  &~~~~~~~~0.&0104528014194     \\
   $c_1$     &&&&  &~~~~~~~24.&51180526856~~      \\
   $c_2$      &&&& &~~~~~~295.&1283580927~~~      \\
   $c_3$      &&&& &~~~~~~~41.&89721595187~~      \\
   $c_4$     &&&&  &~~~~1,168.&592235964~~~~       \\
   $c_5$     &&&&  & ~~~~~111.&6330356226~~~        \\
   $c_6$     &&&&  &~~~~~~~~0.&3315320354509        \\
   $c_7$      &&&& &~~~~3,145.&824919010~~~~   \\
\enddata
\tablenotetext{a}{
 $L(\phi) = {{1 - \phi (c_1 - \phi c_2 )} \over 
 {1 - \phi (c_3 - \phi c_4 ) }}
 e^{-c_5 {\phi}^2} (1 - c_6 e^{-c_7 (\phi - \psi)^2})$
 }
\tablenotetext{b}{$MPP = a L(\phi - \phi_0) + b;~~0. \le \phi \le 1.0$}
\end{deluxetable}

\begin{deluxetable}{rlllccrl}
\rotate
\tablewidth{0.0pt}
\tablecaption{Ephemerides for \psr}
\tablehead{
\colhead{\#} & \colhead{Epoch\tablenotemark{a}} & \colhead{$\nu$} & \colhead{$\dot\nu$}
&\colhead{$n$} & \colhead{$\ddot\nu$}& \# fit & Reduced \\
 & \colhead{(MJD)} & \colhead{(Hz)} & \colhead{$(10^{-10} Hz~s^{-1})$} & &
   \colhead{$(10^{-21} Hz~s^{-2})$} && ~~~$\chi^2$\\ 
       }
 \startdata
   1& 51197.153883667507(~14)&62.040931952(  9)&-1.992409(~17)&  ~3.5\tablenotemark{b}~~~~
   &   2.2~~~~~     &11&~~4.88\\
   2& 51423.703547191827(~20)&62.037074298(  2)&-1.992233(  3)&   6.9(  1)&   4.4(  1)&31&~~4.20\\
   3& 51628.894015752653(~35)&62.033570178(  8)&-1.992751( 17)&  12.3( 14)&   7.9(  9)&17&~~2.16\\
   4& 51738.866765475411(~89)&62.031696343( 30)&-1.993109( 59)&  18.4( 46)&  11.8( 29)&18&~~1.60\\
   5& 51854.122714277637(~~9)&62.029720615(  9)&-1.992993(~75)&  ~3.5\tablenotemark{b}~~~~
   &   2.2~~~~~     & 8&~~0.17\\
   6& 51911.259613446900(~~9)&62.028745439(  4)&-1.992928(~20)&  ~3.5\tablenotemark{b}~~~~
   &   2.2~~~~~     & 9&~~3.25\\
   7& 51994.413363674373(162)&62.027341881( 28)&-1.993125( 28)&  10.6( 11)&   6.8(  7)&25&~~2.37\\
   8& 52209.375211566255(~~9)&62.023652330(  4)&-1.993240(~21)&  ~3.5\tablenotemark{b}~~~~
   & 2.2~~~~~     & 9&~~1.26\\
   9& 52260.402958189564(305)&62.022799990( 69)&-1.993398( 92)&  10.6( 48)&  ~6.8( 31)&22&~~0.70\\
  10& 52415.442549541617( 10)&62.020140505(  3)&-1.993352(~18)&  31.5\tablenotemark{b}~~~~
  &  20.2~~~~~~    &11&~~1.06\\
  11& 52460.230239222541( 13)&62.019382706(  6)&-1.993317(~10)&  ~3.5\tablenotemark{b}~~~~
  &   2.2~~~~~     &13&~~1.80\\
  12& 52570.945740823919(194)&62.017502072( 34)&-1.993557( 33)&  11.0( 13)&   7.1(  8)&27&   ~~1.37\\
  13& 52768.096830064879( 11)&62.014115960(  5)&-1.993603(~32)&  ~3.5\tablenotemark{b}~~~~
  &   2.2~~~~~     & 8&~~0.32\\
  14& 52823.910323497775( 11)&62.013170456(  8)&-1.994199(~15)&  35.0\tablenotemark{b}~~~~
  &  22.4~~~~~~    &12&~~1.29\\
  15& 52930.878000708089( 51)&62.011342395( 13)&-1.993621( 28)&  19.6( 22)&  12.6( 14)&20&~~1.06\\
  16& 53038.547433342749( 89)&62.009509017( 34)&-1.993925( 62)&  15.0( 42)&  ~9.6( 27)&14&~~1.35\\
  17& 53140.932753772596( 16)&62.007746465( 63)&-1.994542(641)&  ~3.5\tablenotemark{b}~~~~
  &   2.2~~~~~     & 5&~~0.48\\
  18& 53155.873880007801(277)&62.007513323( 65)&-1.994108( 71)&  13.2( 32)&  ~8.5( 20)&15&~~2.21\\
\\
\\
\\
  19& 53312.574046533026(163)&62.004838505( 39)&-1.994275( 44)&  16.0( 19)&  10.2( 12)&20&~~3.44\\
  20& 53450.086572363712( 17)&62.002485827(  6)&-1.994458(~~8)&  19.0\tablenotemark{b}~~~~
  &  12.2~~~~~~    &11&~~0.84\\
  21& 53557.044381976239(573)&62.000663106(128)&-1.994517(133)&  17.5( 58)&  11.2( 37)&16&~~1.49\\
  22& 53812.224185839386(233)&61.996292178( 70)&-1.993782(100)&  12.7( 58)&  ~8.1( 37)&14&~~0.60\\
  23& 53881.096123033291( 13)&61.995120229(  6)&-1.994544( 17)&  15.0\tablenotemark{b}~~~~
  &   9.6~~~~~    &10&~~2.56\\
  24& 53952.687378384607( 26)&61.993888026( 61)&-1.995140(443)&  15.0\tablenotemark{b}~~~~
  &   9.6~~~~~    & 4&~~0.00\\

 \enddata
 \tablenotetext{a}{Arrival time of pulse peak in TDB.}
 \tablenotetext{b}{Assumed value.}
 \end{deluxetable}

\begin{deluxetable}{clrrrrrrcrr}
\rotate
\tablewidth{0.0pt}
\tablecaption{Log of \psr\ Glitches}
\tablehead{
\\
\# & Time &  $\Delta\nu$  & $\Delta\nu/\nu$ & $\delta(\Delta\nu/\nu)$\tablenotemark{a} 
&$\Delta$T$_{after}$ &Corr.\tablenotemark{b} &$\Delta$T$_{scatter}$ & $\Delta\dot\nu$ &
$\Delta\dot\nu/\dot\nu$ & 
   $\delta(\Delta\dot\nu)/\dot\nu$\tablenotemark{a}\\
  & (MJD)       &  ($\mu$Hz) & (ppm) & (ppm) & (days) & (days) & (days) & ($10^{-14}$Hz s$^{-1}$) & (ppm) & (ppm)\\
}
 \startdata
  1&$51,285.7 (8.6)$ &    42.2& 0.681&0.065&  283.3& -14.5 & 268.8 & -8.4
  & 422 &  622   \\
  2&$51,569.0 (6.8)$ &    27.8& 0.449&0.008&  142.1&   0.0 & 142.1 & -14.9& 746 &   49   \\
  3&$51,711.1 (6.7)$ &    19.5& 0.315&0.009&  115.2&   0.0 & 115.2 & -12.1& 606 &   74   \\
  4&$51,826.3 (7.1)$ &     8.7& 0.140&0.007&   55.0& -11.3 &  43.7 & -7.9& 397 &   67   \\
  5&$51,881.3 (5.5)$ &     8.7& 0.141&0.020&   78.8& -12.0 & 66.8 & -3.8& 188 &  326   \\
  6&$51,960.1 (4.9)$ &    28.3& 0.456&0.046&  210.6& -14.0 & 196.6 &-10.6& 532 &  653   \\
  7&$52,170.6 (8.3)$ &    11.5& 0.185&0.006&   71.0&   0.0 & 71.0 & -12.8& 642 &   25   \\
  8&$52,241.6 (7.8)$ &    26.5& 0.427&0.006&  144.4&   0.0 & 144.4 & -3.4& 170 &   44   \\
  9&$52,386.0 (5.7)$ &    10.4& 0.168&0.020&   67.3&   0.0 &  67.3 & -12.1& 607 &  242   \\
 10&$52,453.3 (6.9)$ &    13.5& 0.217&0.030&   92.5&   0.0 &  92.5 & -6.8& 342 &  355   \\
 11&$52,545.3 (6.2)$ &    26.1& 0.421&0.018&  194.0& -22.6 & 171.4 & -7.1& 357 &  175   \\
 12&$52,739.8 (5.3)$ &     9.0& 0.144&0.006&   79.2& -20.5 &  58.7 & -11.2& 564 &   29   \\
 13&$52,819.0 (3.6)$ &    15.9& 0.256&0.016&  67.9&   0.0 & 67.9 & -8.0& 401 &  212   \\
 14&$52,886.9 (4.5)$ &    14.5& 0.234&0.023&  127.1&   0.0 & 127.1 &-11.2& 562 &  263   \\
 15&$53,014.0 (9.5)$ &    21.0& 0.338&0.010&  111.4&   0.0 & 111.4 &-14.1& 708 &   78   \\
 16&$53,125.4 (2.8)$ &     1.1& 0.018&0.014&   19.8&   0.0 & 19.8 &-12.4& 620 &  243   \\
 17&$53,145.2 (2.7)$ &    24.3& 0.392&0.008&  ~143.1&   0.0 & 143.1 & 2.05&-102 &  230   \\
 18&$53,288.3 (2.4)$ &    24.5& 0.395&0.010&  157.3&  -3.2 & 154.1 &-13.5& 677 &   58   \\
 19&$53,445.6 (1.7)$ &    16.1& 0.259&0.016&  105.2&  -10.4 &  94.8 &-14.1& 705 &  125   \\
\\
\\
 20&$53,550.8 (4.4)$ &    19.9& 0.322&0.026&  156.4&  -22.8 & 133.6 & -11.8& 590 &  205   \\
 21&$53,699.2 (3.9)$\tablenotemark{c} &    24.9& 0.402&0.008& 161.4& -17.9& 143.5 & -12.8 
 & 640 &  173   \\
 22&$53,860.1 (1.5)$ &    14.6& 0.236&0.020&   91.1&  -14.6 &   76.5 & -12.5& 625 &  183   \\    
 23&$53,951.2 (1.5)$ &     1.1& 0.018&0.020&$>$17.0\tablenotemark{d}&  ----- &$>$17.0\tablenotemark{d} 
 & -----& --- &  ---   \\    
Average && 17.8 & 0.288 && 121.6 & -7.1 &  114.1 & -10.0 & 500 \\

 \enddata
 \tablenotetext{a}{These errors come from the formal errors on $\nu$ and
 $\dot\nu$ of the polynomial fits.  
    }
 \tablenotetext{b}{The correction in days needed to convert the time to 
 the next glitch into the time to the beginning of quasi-stable behavior}
 \tablenotetext{c}{This epoch is taken before the last unphased point at 
 MJD 53,702.  Rather than attempting to separate the highly uncertain $\Delta\nu$ 
 contribution of this point (2.46 $\pm$ 3.28 $\mu$Hz) to glitch 21,
 we use this earlier epoch and the full, more accurate $\Delta\nu$ of 25.22 
 $\mu$Hz across the entire inter-timing solution gap (MJD 53,681.6 to 53,710.6).  
 In this case we need only subtract the the microglitch at the end of data 
 segment 21, which was also accurately determined to be 0.29 $\mu$Hz, leaving 
 24.93 $\mu$Hz. }
 \tablenotetext{d}{Not included in the average listed at the bottom}
\end{deluxetable}

\begin{deluxetable}{lclrrrrrrr}
\rotate
\tablewidth{0.0pt}
\tablecaption{Log of Large Glitches from other Pulsars}
\tablehead{
 \\
 PSR J & Time &  ~~~~$\nu$  & $\Delta\nu$~~~ & $\Delta\nu / \nu$~~ & $\dot\nu~~~~$ 
 & -$\Delta\dot\nu$~~~  & $\Delta\dot\nu / \dot\nu$ & Q~~~   &  Ref.\tablenotemark{a} \\
&(MJD)&~~(Hz)&($\mu$Hz)~~&(ppm)~~~&(fHz/s)&(fHz/s)& ($10^{-3}$) & ($10^{-3}$)       \\
}
 \startdata
 0205+6449&52,555(13)&15.22230&15.2(1.5)&1.0~(0.1)&-44956.~& $----$    & $----$  & $----$    & ~~~~1 \\
 0358+5413&46,497(~4)&~6.39463&27.93( 1)&4.638(~1)&~~-179.8&~12.6(1.8)&70(10)~~&  1.12(7) & ~~~~2\\
 0835-4510&40,280(~4)&11.20978&26.23(11)&2.340(10)&-16326.~&115.1~~~~~&~7.08\tablenotemark{b}~~
          & 30(10)   & ~~3,4 \\
          &41,192(~8)&11.20855&22.98(34)&2.050(30)&-16257.~&117.9~~~~~&~7.25\tablenotemark{b}~~
	  & 35.(1)   & ~~3,4 \\
          &42,683(~3)&11.20654&22.30(11)&1.990(10)&-16145.~&114.~~~~~~&~7.1\tablenotemark{b}~~~
	  & 88.(8)   & 3,4,5 \\
          &43,693(12)&11.20518&34.29(67)&3.060(60)&-16069.~&104.1~~~~~&~6.48\tablenotemark{b}~~
	  & 24.(5)   & 3,4,5 \\
          &44,888(~0)&11.20357&12.83( 3)&1.145(~3)&-15980.~&102.~~~~~ &~6.39\tablenotemark{b}~~
	  &183.(1)   & ~~3,6 \\
          &45,192(~1)&11.20316&22.97(11)&2.050(10)&-15956.~&~97.~~~~~ &~6.1\tablenotemark{b}~~~
	  & 44.(3)   & ~~3,7 \\
          &46,257(~0)&11.20173&17.93(11)&1.601(~1)&-15875.~&103.~~~~~~&~6.5\tablenotemark{b}~~~
	  &158.(1)   & ~~~~7 \\
          &47,520(~0)&11.20000&20.24( 1)&1.807~(1)&-15780.~&~68.2~~~~~&~4.32\tablenotemark{b}~~
	  &$----$    & ~~8,9 \\
          &48,457(~0)&11.19877&30.40( 2)&2.715~(2)&-15709.~& $----$   & $----$  &$----$   & ~~~10 \\
          &49,559(~0)&11.19729&~9.35( 2)&0.835~(2)&-15626.~& $----$   & $----$  &$----$   & ~~~11 \\
          &50,369(~0)&11.19620&23.62(13)&2.110(12)&~-1556.5&~92.6(5)~~&~5.95(3)&380(2)    & ~~~12\\
          &51,559(~0)&11.19462&34.54(~0)&3.086(~0)&~-1556.1&104.8(0)~~&  $----$ &$----$   & ~~~13 \\
 1048-5832&49,034(~9)&~8.08700&24.22( 4)&2.995(~7)&~-6235.1&~23.1(6)~~&~3.7(1)~&  2.5(3)  & ~~~14  \\
          &50,788(~3)&~8.08604&~6.22( 6)&0.771(~7)&~-6236.0&~28.8(4)~~&~4.62(6)&$----$    & ~~~14 \\
\\
\\
\\
\\
 1341-6220&47,989(10)&~5.17376&~7.78( 4)&1.505(~8)&~-6774.3&~~1.0(4)~~&~0.15(6)& $\le$1   & ~~~15  \\
          &48,645(~5)&~5.17352&~5.12( 4)&0.990(~3)&~-6775.9&~~4.7(7)~~&~0.7(1)~&  2.0(3)  & ~~~15  \\
          &50,008(16)&~5.17274&~8.46( 7)&1.636(13)&~-6771.5&~~2.2(3)~~&~3.3(4)~&  4       & ~~~15  \\
 1357-6429&52,012(16)&~6.02118&14.60(10)&2.420(20)&~-1307.4&~70.2(3)~~&~5.37(2)& $----$   & ~~~16  \\
 1539-5626&48,465(10)&~4.1086 &11.47( 8)&2.791(~2)&~~~-81.9&~~4.~(1.)~&~0.4(1)~& $\le$1   & ~~~17 \\
 1614-5048&49,803(16)&~4.31688&27.87(28)&6.456(65)&~-9241.8&~89.6(1.8)&~9.7(2)~&538(11)   & ~~~14  \\
 1709-4429&48,780(15)&~9.76095&19.64( 8)&2.0123(2)&~-8891.7&~~1.8(5)~~&~0.20(6)&133(7)    & ~~~17 \\
 1730-3350&47,990(10)&~7.1714 &22.10( 7)&3.080(10)&~-4643.4&~56.~(14)~&12.(3)~~&  8(1)    & ~~~17 \\
 1801-2451&49,476(~6)&~8.00694&16.00( 6)&1.998(~7)&~-8172.6&~39.2(~3)~&~4.8(~3)&188(12)   & ~~~14,18 \\
          &50,651(~5)&~8.00612&~9.90( 3)&1.237(~4)&~-8206.9&~31.7(~7)~&~3.87(9)&202(~6)   & ~~~14  \\
 1803-2137&48,245(10)&~7.4849 &30.50(11)&4.075(15)&~-7536.2&~69.3(~3)~&~9.2(~4)& 15(~2)   & ~~~~2 \\
          &50,765(15)&~7.48326&23.95(20)&3.200(27)&~-7488.9&~79.7(11)~&10.7(15)&161(~6)   & ~~~14  \\
 1806-2125&51,062(242)&~2.0756 &32.41(02)&15.615(15)&~~-505.9&~17.(~2)~~&33.6(40)&$----$    & ~~~19  \\
 1826-1334&46,507(40)&~9.85827&26.79(49)&2.718(26)&~-7275.6&~72.~(~7)~&~6.8(1)~&$----$    & ~~~~2  \\
          &49,014(40)&~9.8567 &30.16(49)&3.060(50)&~-7235.4&~72.~(~7)~&10(1)~~~& 18(1)    & ~~~~2  \\
 1833-0827&48,041(10)&11.7258 &21.87( 2)&1.8648(3)&~-1260.6&~~1.93(6)~&~1.53(4)&  0.8(2)  & ~~~~2 \\
\\
\\
\\
\\
 1932+2220&50,264(10)&~6.92221&30.85( 5)&4.457 (6)&~-2756.4&~~4.7 (5)~&~1.7~(2)& $----$    & ~~~20 \\
 2021+3651&52,630(~0)&~9.64114&24.94( 2)&2.587 (2)&~-8886.5&~55.~~(3)~&~6.2~(5)& $----$    & ~~~21  \\
 \enddata
 \tablenotetext{a}{Refs: 1. Ransom et al.~(2004); 2. Shemar \& Lyne (1996); 
 3. Cordes, Downs, \& Krause-Polstorff (1988); 4. Downs (1981); 5 Manchester et al.~(1983); 
 6. McCulloch et al.~(1983); 7. McCulloch et al.~(1987); 8. Flanagan (1990);
 9. McCulloch et al.~(1990); 10. Flanagan (1991); 11. Flanagan (1994a); 12. Flanagan (1994b);
 13. Dodson, McCulloch, \& Lewis (2002); 14. Wang et al.~(2000); 15. Kaspi et al.~(1992)
 16. Camilo et al.~(2004); 17. Johnston et al.~(1995); 18. Arzoumanian et al.~(1994);
 19. Hobbs et al.~(2002); 20. Krawczyk, et al.~(2003); 21. Hessels et al.~(2004).}
 \tablenotetext{b}{These values are the part of the spindown jump associated
 with the long term recovery and are taken from Alpar et al.~(1993).}
\end{deluxetable}
 
\begin{figure}
\vskip 7 in
\includegraphics{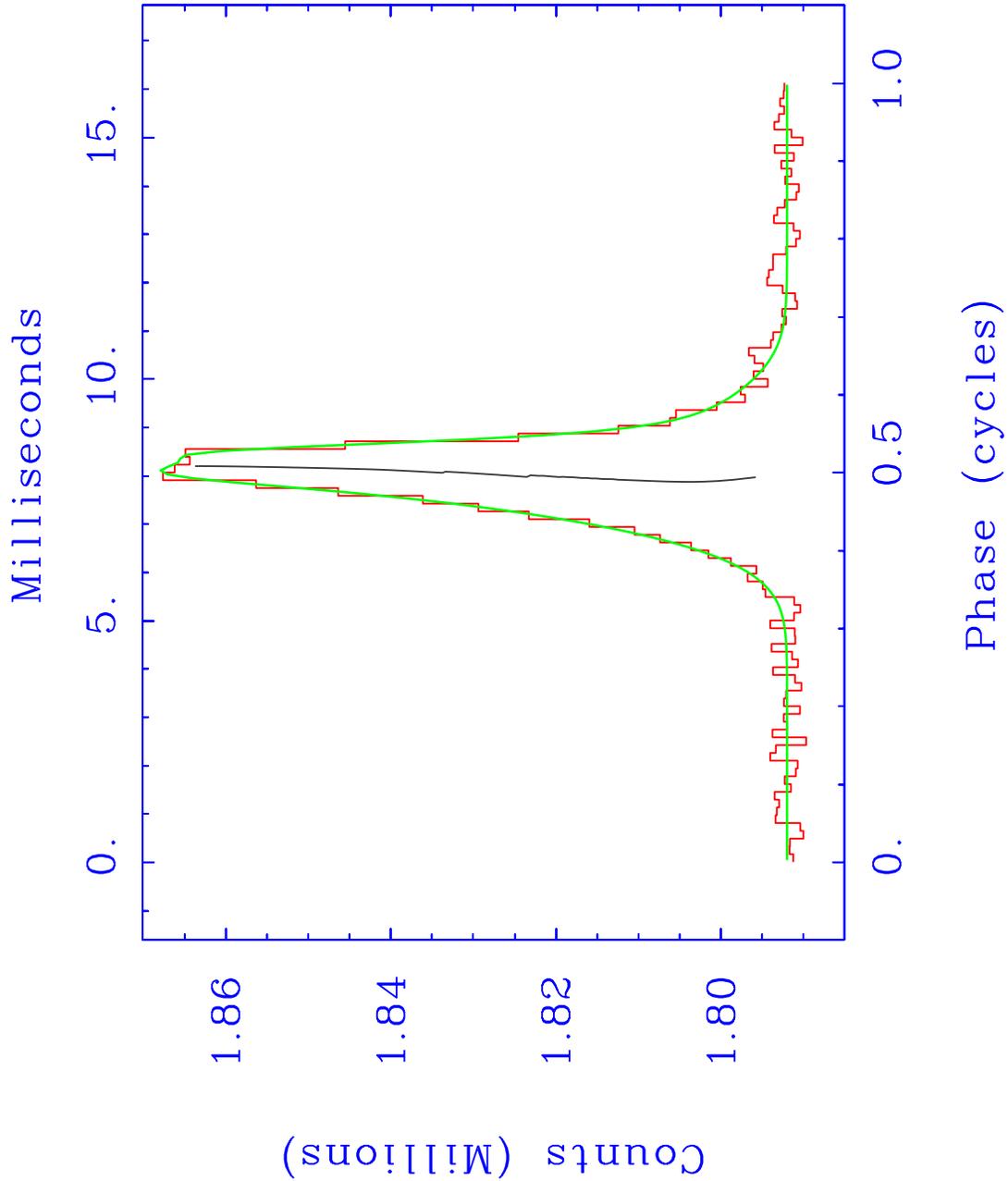}
\figcaption{The pulse profile for \psr\ from 311 observations taken 
during the  RXTE campaign from Jan.~19, 1999, to Oct.~6, 2004.  The 
parameters of the MPP curve drawn over the summed data are listed in Table 2.
The vertical curve marks the midpoint of the pulse profile vs.~height.
     }
\label{fig:PP}
\end{figure}

\begin{figure}
\vskip 7 in
\includegraphics{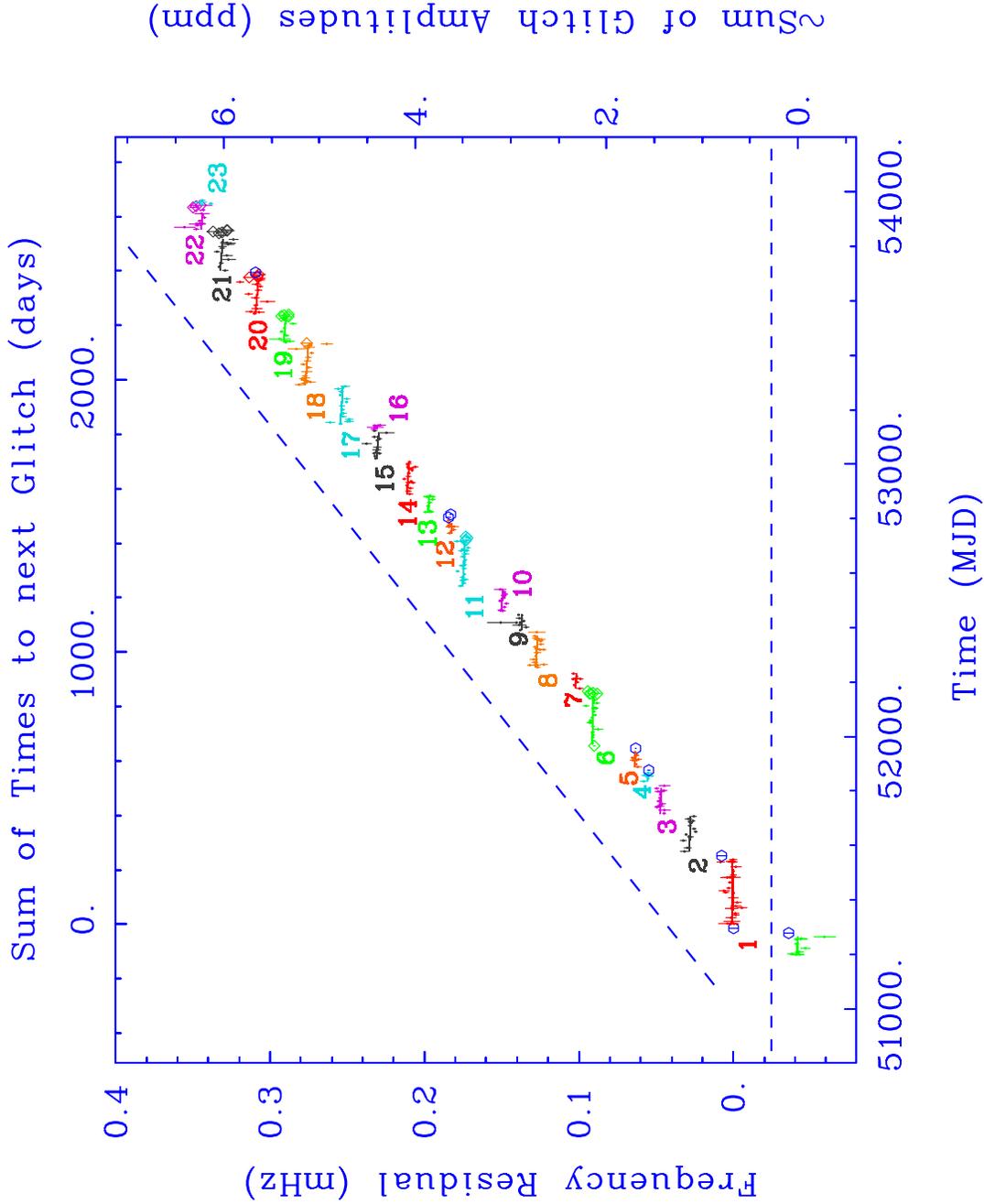}
\figcaption{The frequency history for \psr.  The frequency measurements
of the individual observations are shown with error bars.  The (mostly
horizontal) curves are the frequency histories derived from the much more
accurate timing solutions.  The right frame boundary labels
the {\it approximate} cumulative gain of frequency across the glitches
(see Fig.~8).
Diamonds mark points with known phases from the timing solutions,
but which were not included in them.  Hexagons mark observations
for which no timing solution exists.  
The horizontal dashed line is the baseline for the
frequency residuals, and corresponds to 62.037465 $-$ 
1.9922$\times $10$^{-10}\times$(MJD $-$ 51,400) 
$\times$86400 Hz.
The oblique dashed line is an approximation to the
mean trend for the frequency residuals,
and corresponds to 62.037545987 $-$ 
1.9760$\times $10$^{-10}\times$(MJD $-$ 51,400)$\times$86400 Hz.
     }
\label{fig:allfreq}
\end{figure}

\begin{figure}
\vskip 7 in
\includegraphics{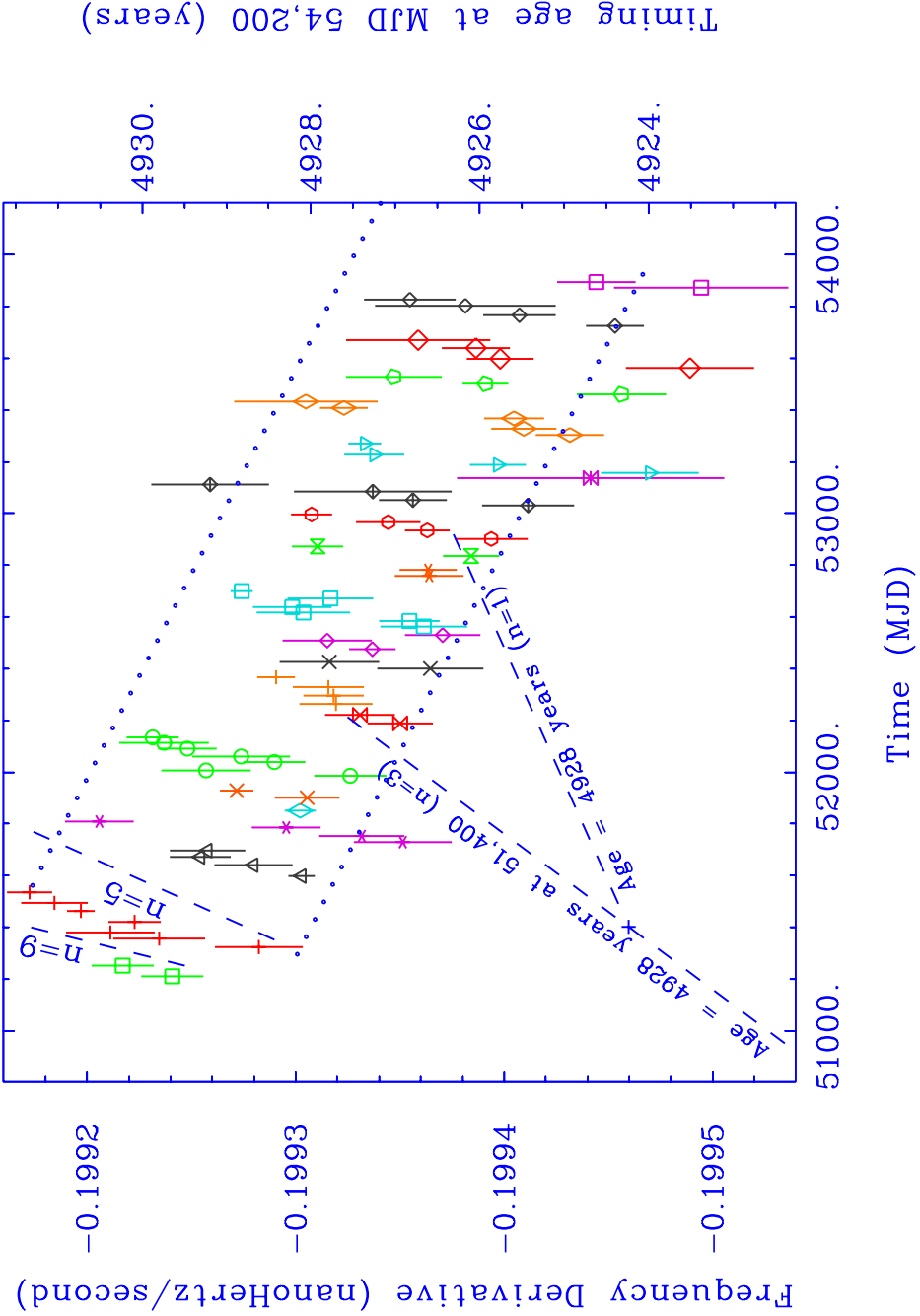}
\figcaption{The $\dot\nu$ or spindown history for \psr, derived from
subsegments of at least three consecutive observations with a time span 
of at least 1,400,000 s (with the exception of the 17th data segment,
which only spanned 14 days), and at least one observation within 0.35 of the 
full time span of its center.  The two dotted lines are spaced 
at 0.15 pHz s$^{-1}$ and slope downward at 0.95$\times$10$^{-21}$ Hz s$^{-2}$,
or 0.15 pHz s$^{-1}$ (five years)$^{-1}$, or a braking index, $n$, of $-$1.5.
     }
\label{fig:allfdot}
\end{figure}

\begin{figure}
\vskip 7 in
\includegraphics{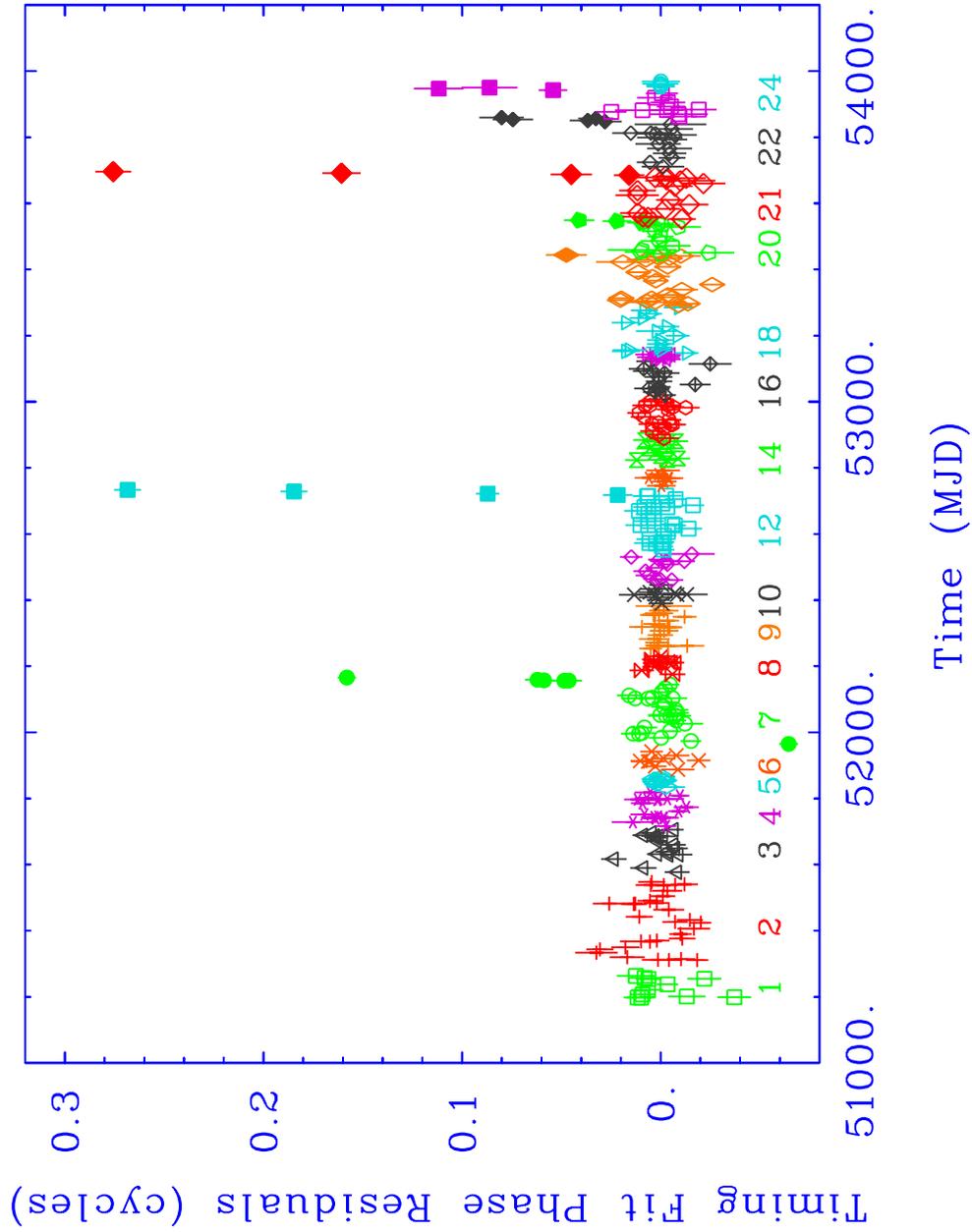}
\figcaption{The phase residuals to the timing solution fits for all 24 
data segments, with the segment numbers listed just below the
corresponding residuals.  The solid points have not been included
in the fits.  The residuals for observations with no timing 
solutions have not been plotted. 
     }
\label{fig:allphase}
\end{figure}

\begin{figure}
\vskip 7 in
\includegraphics{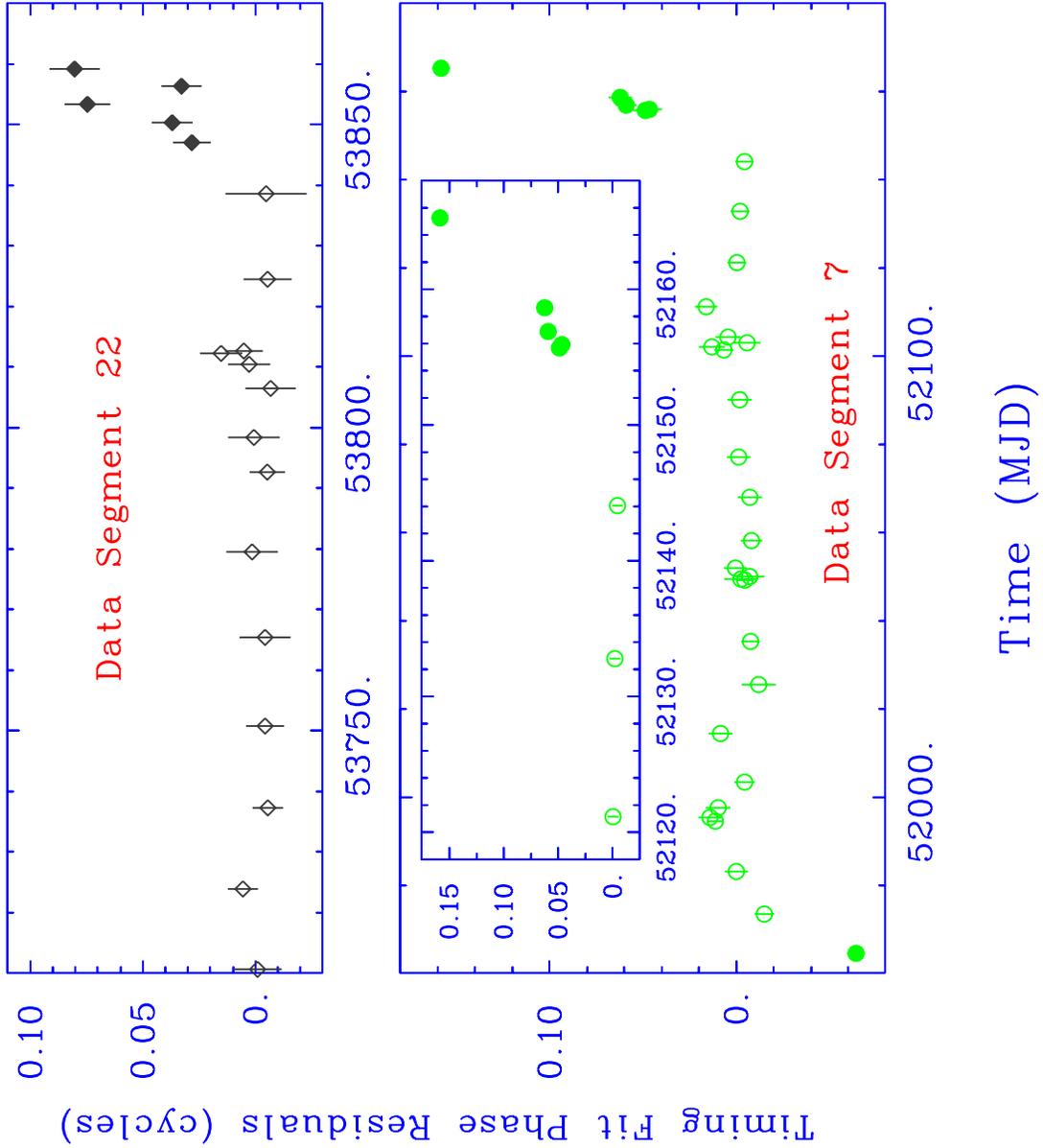}
\figcaption{(Lower) The timing residuals for data segment 7.  The
solid points have not been included in the fit. (Upper) The timing
residuals for data segment 22.
     }
\label{fig:seg7_22phi}
\end{figure}

\begin{figure}
\vskip 7 in
\includegraphics{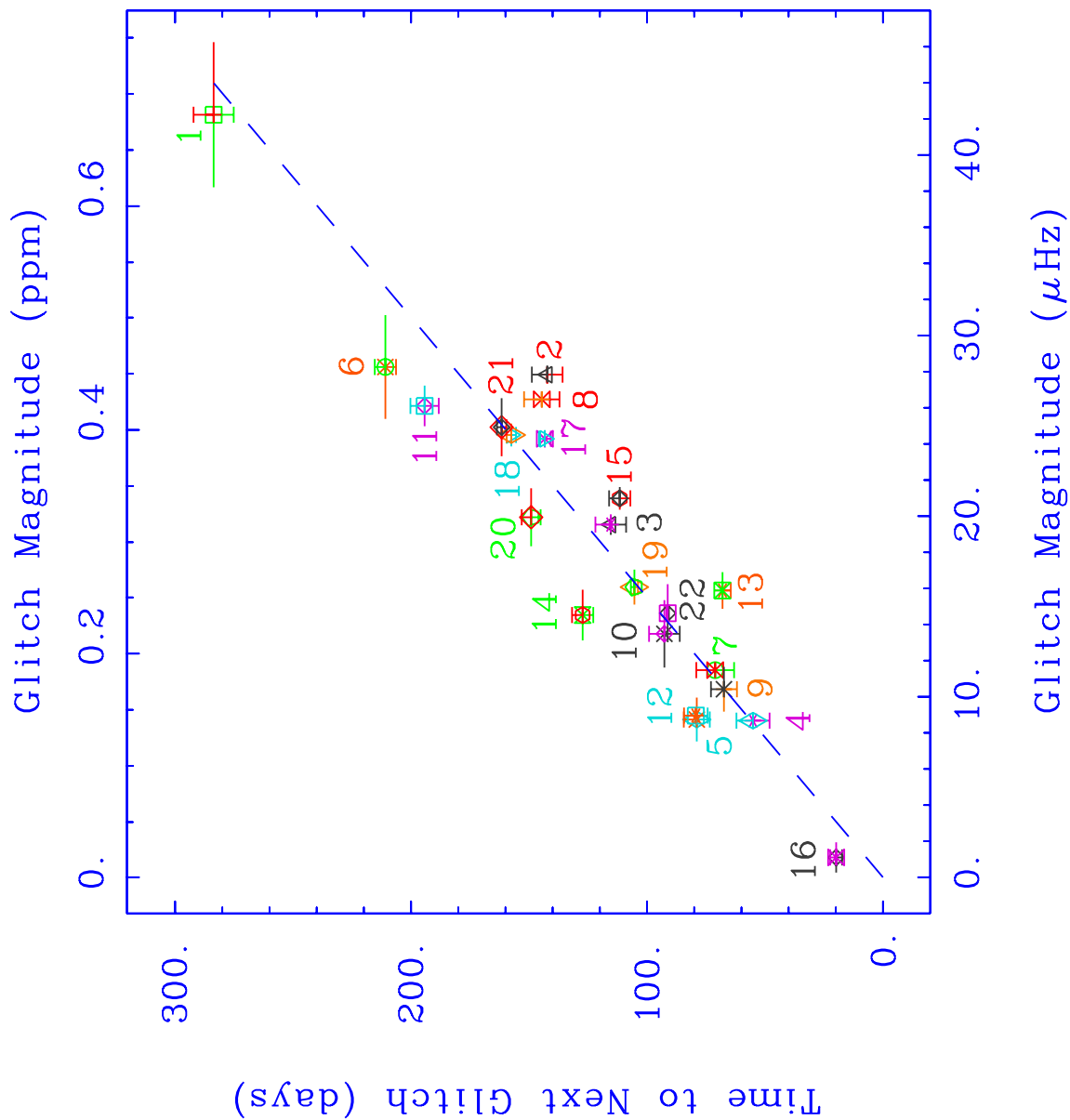}
\figcaption{The glitch amplitude vs.~the time to the next 
glitch (see Table 4).  The time bounds drawn for the points
are equal (up and down) and each is a quarter of the sum of 
two time intervals bounding the data segment.  The slope
of the dashed line fitted to the points and through the
origin is 6.4394 days $\mu$Hz$^{-1}$, or 399.37 days ppm$^{-1}$.
     }
\label{fig:tvsppm}
\end{figure}

\begin{figure}
\vskip 7 in
\includegraphics{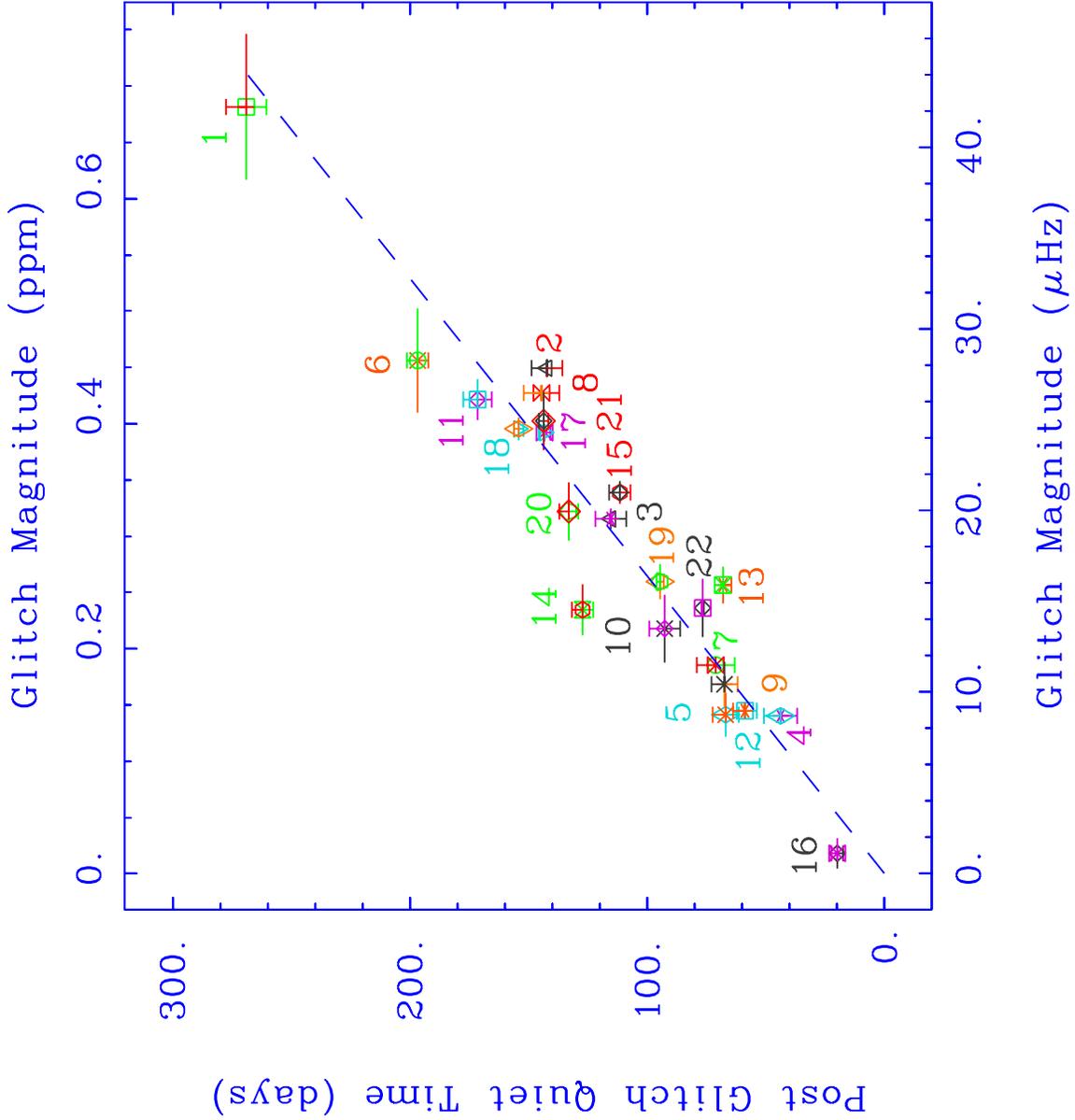}
\figcaption{The glitch amplitude vs.~the following interval 
of stable timing behavior (see Table 4).  The slope of the
dashed line fitted to the points and through the origin is 
6.0955 days $\mu$Hz$^{-1}$, or 378.04 days ppm$^{-1}$.  The point
for glitch 21 is very close to that for glitch 17 (24.9 $\mu$Hz, 
143.5 days, vs 24.3 $\mu$Hz, 143.1 days).
     }
\label{fig:tvsppms}
\end{figure}

\begin{figure}
\vskip 7 in
\includegraphics{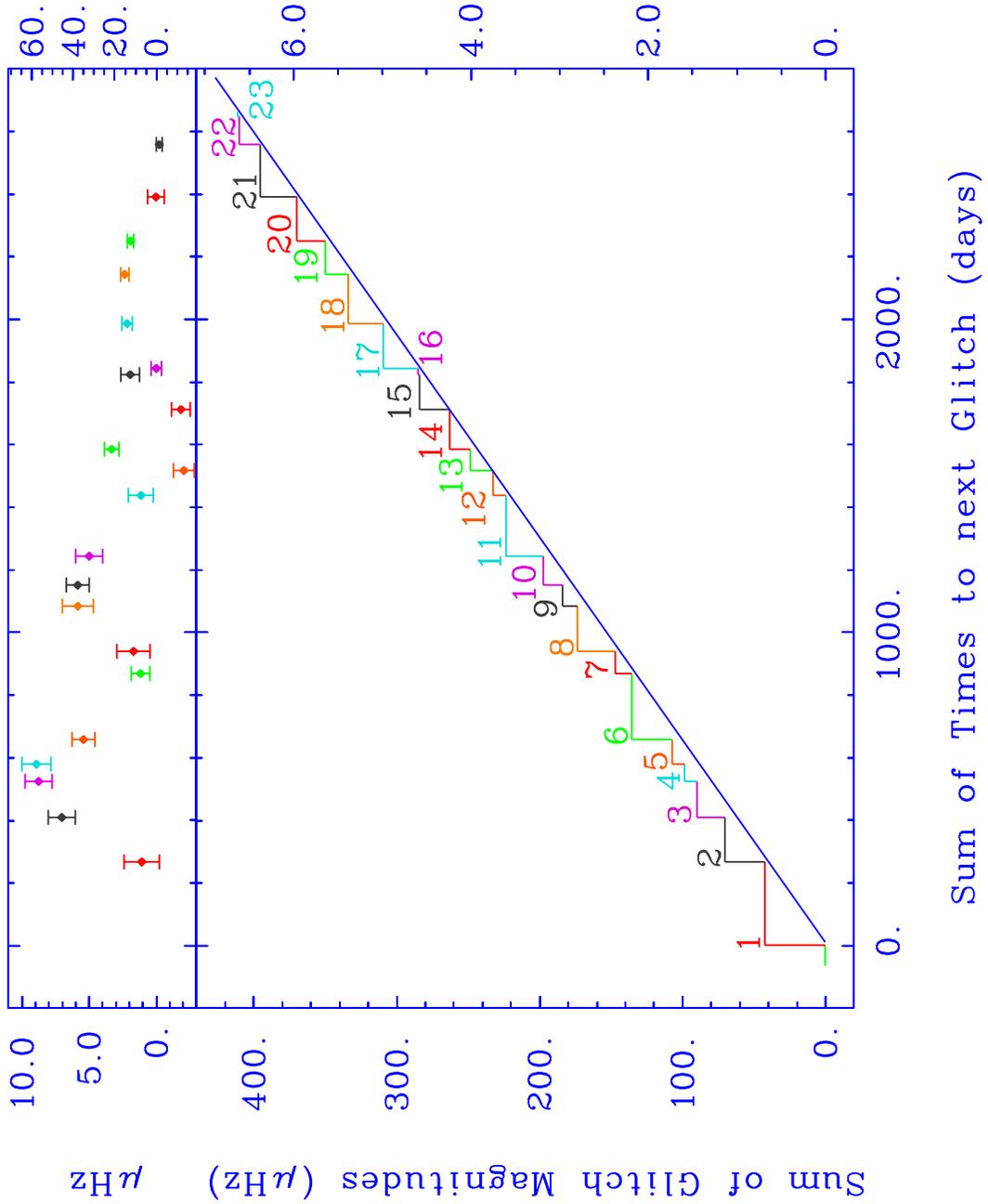}
\figcaption{Cumulative time vs.~cumulative glitch amplitude.
The oblique line drawn from (0,0) to the bottom of glitch 21
has been offset from the first and last glitch times by 
ten days for clarity, and has a slope of 401.1 days ppm$^{-1}$ or
6.468 days $\mu$Hz$^{-1}$.  The top frame plots the predicted
glitch onset time minus the actual glitch onset time (in days
on the right hand scale, and converted to equivalent $\mu$Hz 
on the left hand scale).
     }
\label{fig:cumppmt}
\end{figure}

\begin{figure}
\vskip 7 in
\includegraphics{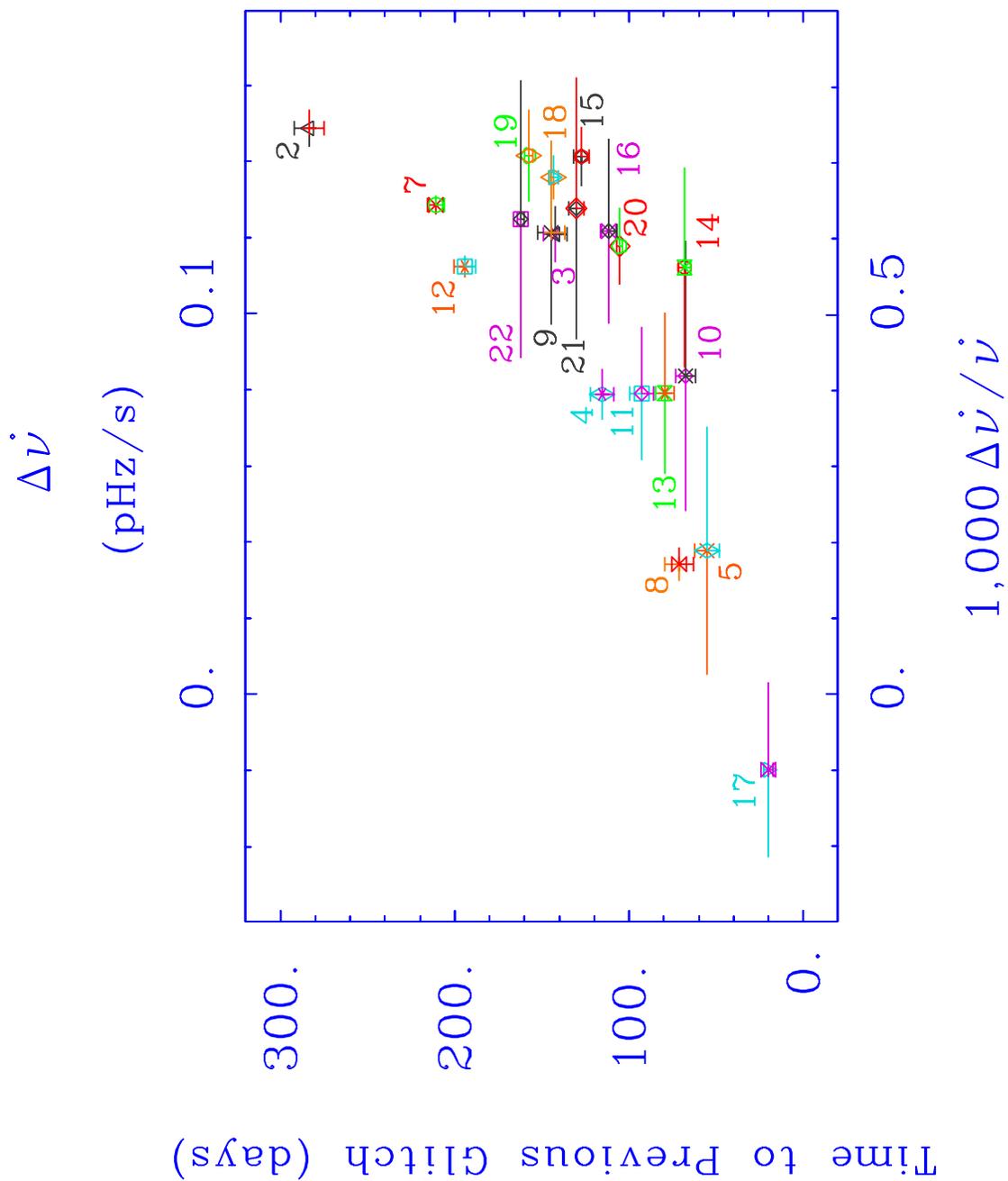}
\figcaption{The time to the previous glitch vs.~$\Delta\dot\nu$ across
the glitch.  
Glitch 6 has been omitted due 
to large errors in $\Delta\dot\nu$ (see Table 4).
     }
\label{fig:tvsfdot}
\end{figure}

\begin{figure}
\vskip 7 in
\includegraphics{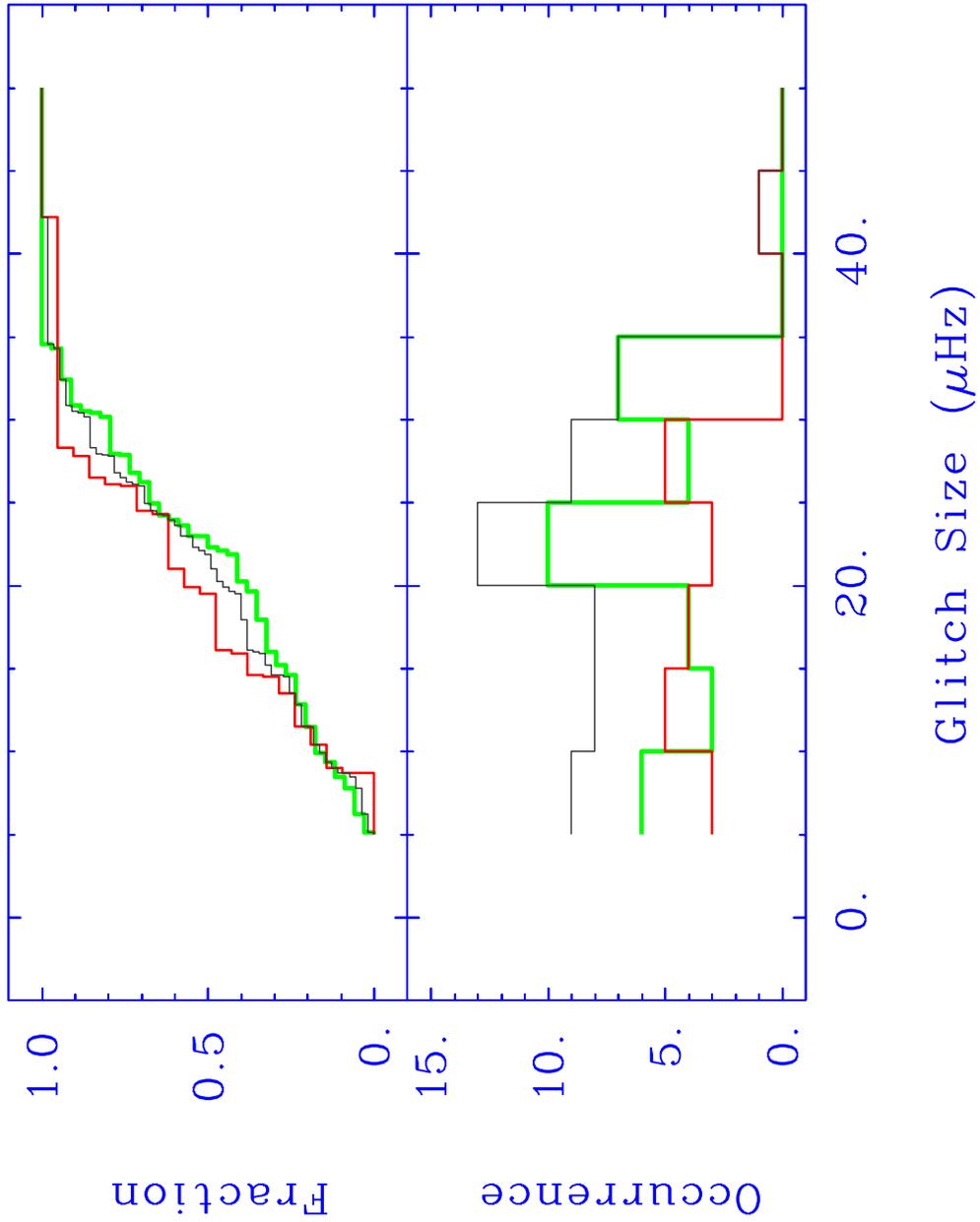}
\figcaption{The distribution in size of 21 large glitches from \psr\
(medium thickness line), and 34 more glitches from other pulsars (thick line), 
all larger than 5 $\mu$Hz, in addition to the combination of both samples
(thin line).
     }
\label{fig:allglitch}
\end{figure}

\begin{figure}
\vskip 7 in
\includegraphics{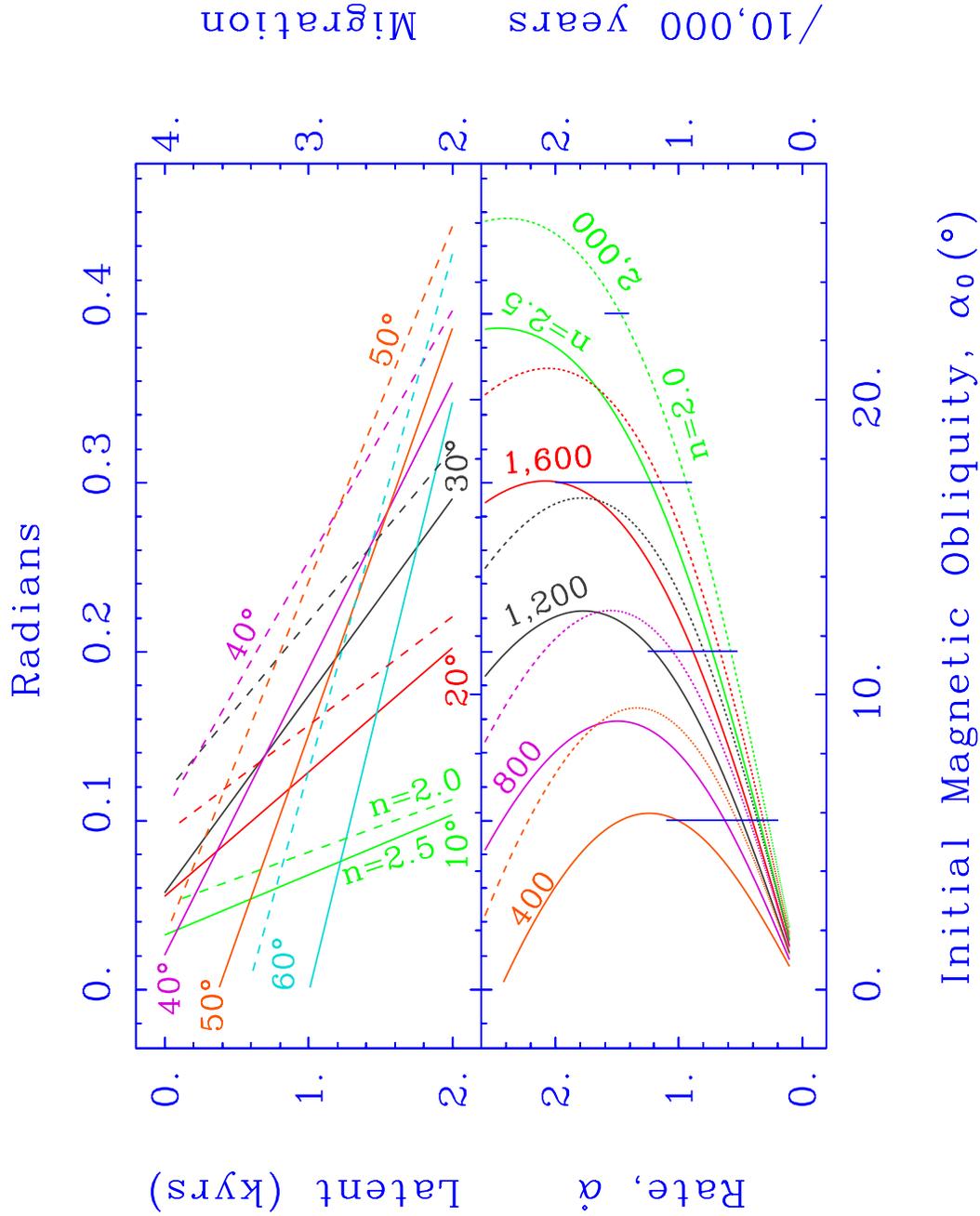}
\figcaption{(Lower) The magnetic pole migration rate vs.~initial obliquity
of \psr\ for an age of 4,000 years, latency times (with no polar 
migration) from 2,000 to 3,600 years in steps of 400 years, and two 
values of the (intrinsic) braking index, $n$, 2.5 (solid) and 
2.0 (dotted/dashed).  The four vertical lines of constant initial
obliquity, or $\alpha_0$, intersect the curves at 22 migration rates, 
or $\dot\alpha$'s, for which time histories of $\nu$ and $\dot\nu$ 
are plotted in Fig.~12 (see $\S$ \ref{sec:Migrate}). (Upper)
Lines of constant present day magnetic obliquity, assuming a 4,000-year
age for \psr, for braking indices 2.5 (solid) and 2.0 (dashed) are
plotted on the initial obliquity-time plane.
See Fig.~12 and $\S$ \ref{sec:Migrate}.
     }
\label{fig:obaltau}
\end{figure}

\begin{figure}
\vskip 7 in
\includegraphics{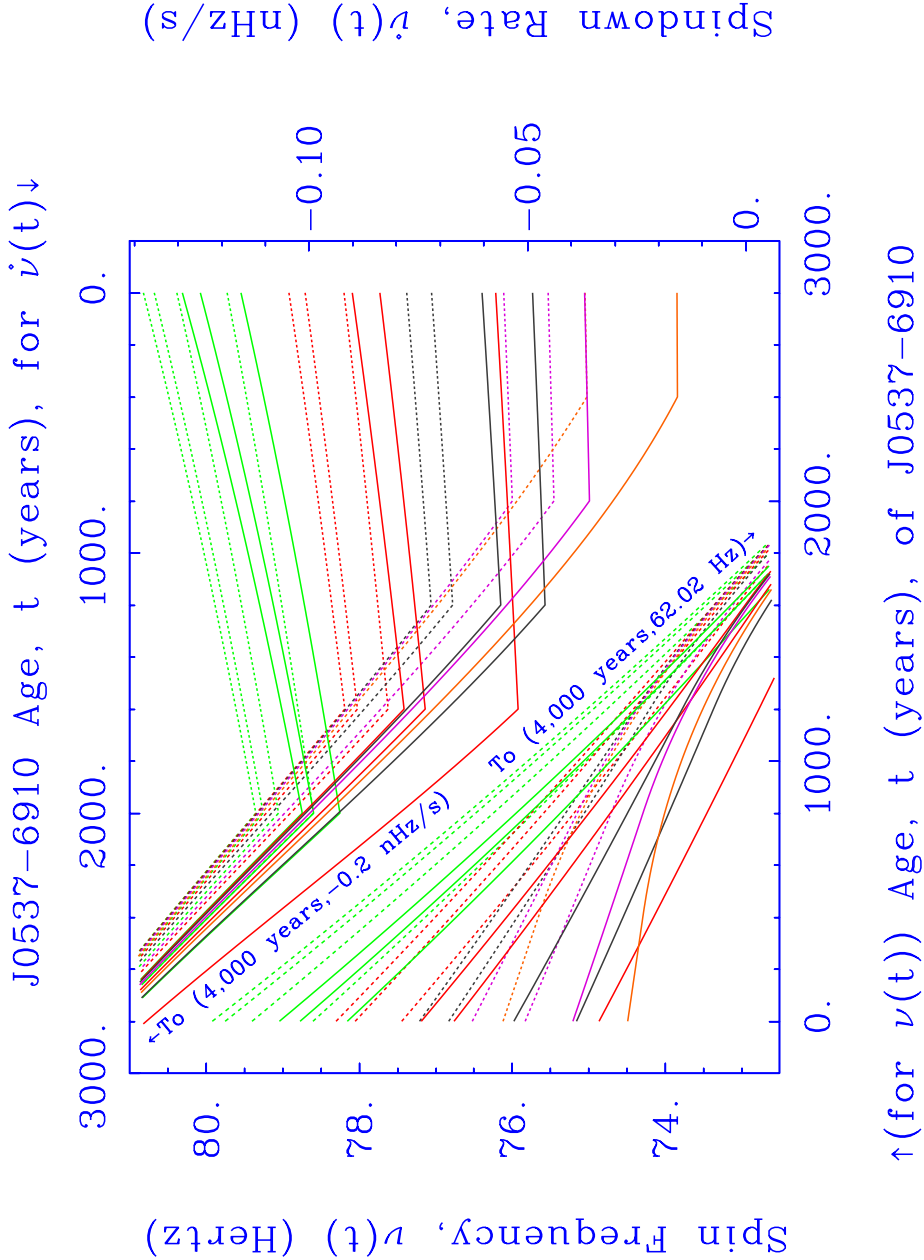}
\figcaption{Segments of 22 possible time histories of \psr\ for
a present age of 4,000 years, and for parameters from the
intersections of ordinates of
Fig.~11 representing initial magnetic obliquities of 0.1 -- 0.4 radians 
in steps of 0.1 radians, with the lower migration rate segments of curves of 
constant pole-migration intervals from 2000 to 3600 years in steps of 400 
years, and braking indices, $n$, of 2.5 (solid) and 2.0 (dotted -- see 
Fig.~11).  The pulse frequency, $\nu$, histories are curves with values 
labeled on the left hand and bottom frame boundaries, first visible at the 
bottom center and continuing left/backward in time, while rising to the 
left-hand frame edge.  The curves with the slope breaks are spindown 
histories and are labeled on the top and right-hand sides.
     }
\label{fig:brake}
\end{figure}

\end{document}